%% file: Boo.tex
\documentstyle[12pt,epsfig,twoside]{article}

\include{definitions}

\epsfverbosetrue
\begin{document}

\newcommand {\badnumber} {685}
\newcommand {\badver} {1}
%

%

\vspace{1.cm}

\begin{flushright}
LAL 04-10
\end{flushright}

\begin{center}
{\LARGE\bf 
Isospin constraints from/on $B \to \pi\pi$
}

\vskip 1.truecm
\centerline{\bf \large M.~Pivk$^{\,a}$ and F.R.~Le~Diberder$^{\,b}$}
\vspace{0.9cm}
{\small \em $^{a}$ CERN, PH Department \\
CH-1211 Geneva 23, Switzerland}\\[0.2cm]

{\small \em $^{b}$ Laboratoire de l'Acc\'el\'erateur Lin\'eaire,\\
                   IN2P3-CNRS et Universit\'e de Paris-Sud, 
                   F-91898 Orsay, France}

\end{center}
\vspace{3cm}

{\abstract{\it The Standard Model constraints on~$\alpha$ which can be derived from the $B\rightarrow\pi\pi$ decays are revisited in some depth. As experimental inputs, the $B^0\rightarrow\pi^+\pi^-$, $B^+\rightarrow\pi^+\pi^0$ decays complemented by the $B^0\rightarrow\pi^0\pi^0$ decays, the CP parameters $\Spipi$ and $\Cpipi$, and/or
the value of~$\alpha$ as determined by the global CKM fit are used.
The constraints discussed here are model independent in the sense that they rely only on Isospin symmetry, following the Gronau-London proposal. A new bound on $\Broo$ and the function $\Coo(\Broo)$ are introduced. While another bound applied to \babar\ results is shown to imply that $\cos(2\alphaeff)$ is negative. The Grossman-Quinn bound is rediscussed. A close form expression is given for~$\alpha$ as a function of the measurements. Various scenarii for the future of the isospin analysis are explored.
To probe the Standard Model the $(\Broo,\Coo)$ plane is introduced.}}

\vspace{5cm}

\newpage
\tableofcontents
\newpage

\vskip .5truecm
\section{Introduction}

The purpose of this note is to revisit the Standard Model constraints which can be derived from (and on) the $B\rightarrow\pi\pi$ decays following the Gronau-London proposal~\cite{bib:GL}. Much have been written already on this topic, for example~\cite{bib:BaBarBook,bib:pipi,bib:ali}, but fresh developments are presented here.

This study leads to correct a strong bias: the $B\rightarrow\pi\pi$ isospin analysis is not doomed neither to wait for ultra high luminosity, nor to have a small branching ratio $\Broo$ (corresponding to the $B^0\rightarrow\pi^0\pi^0$ decay) to be able to constrain~$\alpha$. The current period of time is very exciting: anything can happen in the next three years, including a rather precise determination of~$\alpha$.   

In Section~\ref{sec:babarandbelle} is recalled the present experimental status: the values of the three branching ratios ($\Brpm=\Br(\Bztopipi)$, $\Brpo=\Br(\Bptopippiz)$ and $\Broo$) and the two time-dependent asymmetry parameters ($\Spipi$ and $\Cpipi$). The statistical meaning of the \babar\ and Belle results for $\Spm$ and $\Cpm$ is discussed, considering two different statistical treatments.

In Section~\ref{sec:isospin}, the plague of the mirror solutions is explicited.
Section~\ref{sec:algebrasection} then turns to some algebra.
A new bound on $\Broo$ is presented Eq.(\ref{eq:rebound}),
from which the sign of~$\cos(2\alphaeff)$ is inferred,
a function $\Coo(\Broo)$ is explicited Eq.(\ref{eq:master}),
a close form expression for~$\alpha$ as a function of the measurements is given, and the meaning of the Grossman-Quinn bound on $\mid\alpha-\alphaeff\mid$ is discussed.

In Section~\ref{sec:perspective} the perspectives for measuring~$\alpha$ are pondered. First an exploration of the space of still open possibilities is attempted. Then two scenarii for the high luminosity regime ($10~\invab$) are considered. For the sake of completeness a nasty and a nice scenario are considered, with $500~\invfb$.

The analysis is reconsidered from another point of view in Section~\ref{sec:StandarModelstestsection}. Instead of the usual $\bar\rho$ and $\bar\eta$ variables, the branching ratio $\Broo$ and the asymmetry term $\Coo$ are taken as key variables to probe the Standard Model,
when dealing with $B\rightarrow\pi\pi$ decays.
The present experimental status 
and the high luminosity potential are discussed in this context.

The ElectroWeak Penguin complications are neglected throughout the note, having in mind that,
when accounted for, they should not utterly blur the picture.

The effects of radiative corrections~\cite{bib:sPlot,bib:TheseMu}, which imply a $5$ to $10\%$ increase on the published branching ratios are not taken into account in the present analysis.

\eject

\section{Comparison of \babar\ and Belle results}
\label{sec:babarandbelle}
The two detectors \babar\ and Belle located respectively at the \abf\ \pep2\ and KEK-B measure branching ratios and time-dependent asymmetries for the $B \to \pi\pi$ decays. If the \babar\ values, for the time-dependent asymmetries, are in the physical region, Belle ones are not. A special statistical treatment is then advised to deal with such a situation where a measurement lies beyond the physical boundaries: this is the purpose of this section. First are recalled in Section~\ref{sec:res} the values measured by \babar\ and Belle. The world average (W.A.) ones are given as well~\cite{bib:HFAG}. Section~\ref{sec:naive} discusses the results using the standard statistical treatment. Section~\ref{sec:nonnaive} turns to the boundary effect via a particular treatment.

\subsection{Present experimental status}
\label{sec:res}
The most recent values of \babar\ and Belle are summarized in Table~\ref{tab:inputs}. All branching ratios are quoted in the $10^{-6}$ unit, as in the whole document.
For the world average values of the two time-dependent asymmetries terms $\Spm$ and $\Cpm$, the discrepancy between \babar\ and Belle is not taken into account and the results are merged together at face values. That is to say the raw measurements are taken, without accounting for the fact that physical values should be in the disk $\Spm^2+\Cpm^2\le 1$.
Since Belle results are outside the physical disk, in some instances modified and fictitious results are used instead, where the Belle measurements are forced back onto the unity circle: they are denoted \sBelle\ in the present document.
\begin{table}[ht]
\begin{center}
\begin{tabular}{|c|c|c|c|c|c|c||}
\hline
    & $\Brpm~(10^{-6})$          & $\Brpo~(10^{-6})$          &$\Broo~(10^{-6})$ & $\Cpm$          & $\Spm$   & $\rho$ \\ \hline\hline 
\babar\      & $4.7\ \pm 0.6\ $ & $5.5\ \pm 1.2\ $ &$2.1\ \pm 0.7\ $ & $-0.19\pm 0.20$ & $-0.40\pm 0.22$ & $-0.02$ \\ \hline 
Belle      & $4.4\ \pm 0.7\ $ & $5.0\ \pm 1.3\ $ &$1.7\ \pm 0.6\ $ & $-0.58\pm 0.17$ & $-1.00\pm 0.22$ &$-0.29$  \\ \hline 
\sBelle     & -                 & -                &   -  & $-0.55\pm 0.17$ & $-0.84\pm 0.22$ & - \\ \hline  \hline    
W.A.       & $4.55\pm 0.44$   & $5.18\pm 0.77$   &$1.9\ \pm 0.5\ $& $-0.46\pm 0.13$ & $-0.73\pm 0.16$ & $-0.17$  \\ \hline\hline  
\end{tabular}
\end{center}
\caption{\it Input values to the fits for the branching ratios~\cite{bib:pipiBabar,bib:BApippi0,bib:BApi0pi0,bib:BRBELLE} and for the CP violating asymmetries~\cite{bib:BAasym,bib:CPpipiBelle}.   The notation \sBelle\ refers to the Belle results for the branching ratios, but where the measured values for $\Spm$ and $\Cpm$ are replaced by the values quoted by Belle for $\Spmbest$ and $\Cpmbest$ (Section~\ref{sec:nonnaive}). The world average branching ratios are obtained using in addition CLEO results~\cite{bib:CLEO1}.}
\label{tab:inputs}
\end{table}

\subsection{The unbounded (standard) statistical treatment}
\label{sec:naive}
A proper method to compute a confidence level (CL) associated to a given theoretical hypothesis (\hyp) and an experimental result (\exp) is the following: one defines a test statistics $w(\hyp,\exp)$ which bears the property that the larger its value, the smaller the likelihood of the hypothesis (\hyp) to be true, given the experimental result (\exp). One denotes $f_{\hyp}(w)$ the probability density function of the variable $w$, the hypothesis being assumed to be true. The confidence level is then computed as:
\begin{equation}
\CL(\hyp,\exp)~=~\int_{w(\hyp,\exp)}^{\infty} f_{\hyp}(w)dw ~.
\end{equation}
When the theoretical hypothesis is unbounded, a standard treatment is to identify $w$ with the $\chi^2$ measuring the agreement between theory and experiment. When the theoretical hypothesis is bounded, this standard treatment is still valid. However, it suffers from a drawback: if, because of a statistical fluctuation, the experimental result lies far away from the allowed region, the confidence level might be very small, whatever the hypothesis. The underlying theory is therefore ruled out. The drawback stems from the fact that one may wish to set a confidence level assuming that the underlying theory is true. That is to say, one does not question the validity of the theory as a whole, but one is interested to reach a statement on some of its theoretical parameters.

\subsection{The bounded statistical treatment}
\label{sec:nonnaive}
The standard, $\chi^2$-based, statistical analysis of the $\Spm$ and $\Cpm$ measurements may lead to the exclusion of the whole physical domain $\Spm^2+\Cpm^2\le 1$
if statistical fluctuations bring the measurements far enough outside it.
Because Belle values enter deep into the unphysical region,
the hypothesis of no CP violation 
in the $B\rightarrow\pi\pi$ ($\Spm=\Cpm=0$) is ruled out at $2.48 \times 10^{-5}\%$ CL by her analysis~\cite{bib:CPpipiBelle}.
Here is presented a statistical treatment which by construction cannot rule out the whole physical domain. It incorporates the (mild) assumption that at least Quantum Mechanics is right and hence that $\Spm^2+\Cpm^2\le 1$.

Let us denote:
\begin{itemize}

\item{}  
$\Spmhyp$ and $\Cpmhyp$ , the theoretical values one would like to set a CL on.
They are within the physical domain in the $(\Spm,\Cpm)$ plane,

\item{}  
$\Spmexp$ and $\Cpmexp$ , the measured values.
They can be anywhere in the $(\Spm,\Cpm)$ plane,
    
\item{}  
$\Spmbest$ and $\Cpmbest$ ,     
the theoretical values which are in the best agreement with the above measurements.
They are within the physical domain in the $(\Spm,\Cpm)$ plane.
These values are the ones used to replace $\Spmexp$ and $\Cpmexp$, when considering the \sBelle~results,

\item{}
$\chi^2(\exp|\hyp)$  is the $\chi^2$ one gets when comparing the measured values
      with the hypotheses $\Spmhyp$ and $\Cpmhyp$,
      
\item{}
$\chi^2(\exp|\best)$ is the minimal $\chi^2(\exp|\hyp)$ one gets when exploring the physical domain.
The corresponding values of~$\Spm$ and $\Cpm$ are the above defined $\Spmbest$ and $\Cpmbest$,
      
\item{}      
$w=\chi^2(\exp|\hyp)-\chi^2(\exp|\best)$ is the test statistics used to defined the Confidence Level in the presently advocated treatment.
\end{itemize}
\begin{figure}[t]
\begin{center}
\mbox{\epsfig{file=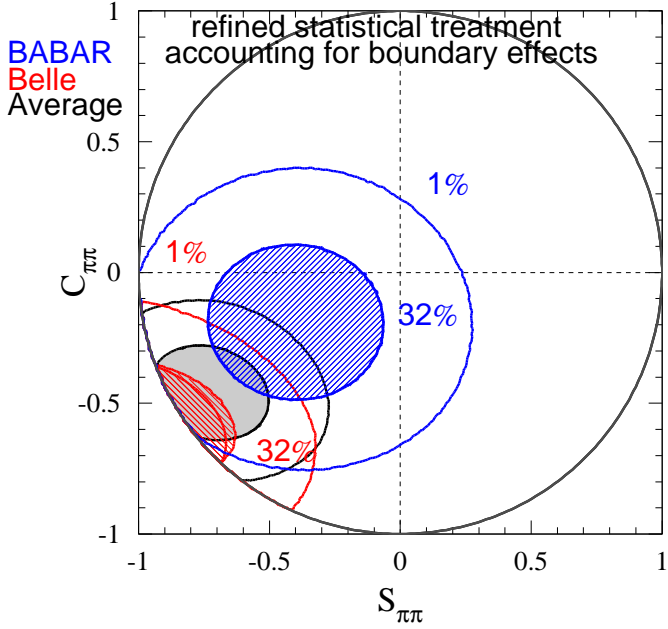,height=9cm}}
\end{center}
\caption{\it Confidence level at $32\%$ and $1\%$ corresponding to \babar\ and Belle results, and their average. The calculation takes into account the fact that the theoretical values are in the physical domain $\Spipi^2+\Cpipi^2 \leq 1$. The ellipse located inside the shaded areas indicate the $32\%$ bound obtained with the usual treatment.}
\label{fig:cl}
\end{figure}
If $\Spmexp$ and $\Cpmexp$ are inside the physical domain then $\chi^2(\exp|\best)=0$.
Whatever $\Spmexp$ and $\Cpmexp$ are, by construction, there is a point in the physical domain where $w=0$: this point is $\{\Spmbest$,$\Cpmbest \}$. It is on the boundary of the physical domain if the measured values are outside~\footnote{The precise location of this point is not at the intersection of the unity-radius circle and the line joigning the measured point and the origin, but not far.}.

For each point $\{\Spmhyp$,$\Cpmhyp\}$, one computes $w$ from the values
of the measurement at hand. Then, one proceeds to a (large) toy 
Monte-Carlo simulation to compute the probability to get a $w$ value larger
than the one above.
In principle, one should first obtain the pdf 
${\cal P}(\Spmexp,\Cpmexp \mid \Spmhyp,\Cpmhyp)$, 
which shape slightly depends on \{$\Spmhyp$,$\Cpmhyp$\}, and Belle does that.    
In practice, a mere Gaussian approximation is used, which is fair enough for the present discussion. Applying this treatment which accounts for boundary effects, one gets the contours of Fig.~\ref{fig:cl} where are drawn the CL computed from \babar\ results (blue), Belle results (red) and from their average. For each of the three above sets of values, three contours are shown: the 32\% (shaded area) and 1\% CL ellipses provided by the treatment, and the 32\% CL (smaller) ellipse obtained with the standard treatment.

\babar\ results are still compatible with no-CP violation, but viewed from Belle or W.A. ones, the hypothesis of no-CP violation is excluded at a very high level.

\section{Isospin Constraints on $B\rightarrow\pi\pi$}
\label{sec:isospin}
The constraints implied by SU(2) on the $B\rightarrow\pi\pi$ decays~\cite{bib:GL} can be expressed as:
\begin{eqnarray}
\Apm&=&\mu\a\ 
\\
\Abarpm&=&\mu\abar\ e^{+2i\alphaeff} 
\\
\Apo&=&\mu\ e^{i(\delta-\alpha_0)}
\\
\Abarpo&=&\mu\ e^{i(\delta+\alpha_0)}
\\
\Aoo&=&\Apo -{\Apm\over\sqrt{2}}=
\mu\ e^{i(\delta-\alpha_0)}\left(1-{\a\over\sqrt{2}}\ e^{+i(\alpha_0-\delta)}\right)
\\
\Abaroo&=&\Abarpo -{\Abarpm\over\sqrt{2}}=
\mu\ e^{i(\delta+\alpha_0)}\left(1-{\abar\over\sqrt{2}}\ e^{-i(\alpha_0+\delta-2\alphaeff)} \right) ~,
\end{eqnarray}
where
\begin{itemize}

\item{}
$\Apm/\Abarpm$ ;$\Apo/\Abarpo$ ;$\Aoo/\Abaroo$ 
are the amplitudes of the $B/\overline{B}\rightarrow\pi^+\pi^-$; $\pi^+\pi^0$,$\pi^-\pi^0$; $\pi^0\pi^0$ decays,
respectively,

\item{}
$\mu$, $\a$ and $\abar$ are three unknown real (and positive) parameters which drive the strenght of the
branching ratios,

\item{}
$\alpha_0$, $\alphaeff$ and $\delta$ are three phases ($\alpha_0$ is the true value of the angle $\alpha$),

\item{}
the phase convention chosen here is such that $\Apm$ is real positive.
\end{itemize}
Although $\alphaeff$ bears information on CP violation,
its value is technically independent of the~$\alpha_0$ value:
the presence of Penguins diagrams breaks down the relationship,
if they are theoretically unknown, which is assumed here. In terms of the above six unknown parameters, the six physical observables take the form~\footnote{To take into account the fact that the lifetime of charged and neutral $B$ are different, every branching ratio should be divided by the corresponding lifetime: $\Brpm$, $\Brpo$ and $\Broo$ become then respectively $\frac{\Brpm}{\tauBo}$, $\frac{\Brpo}{\tauBp}$ and $\frac{\Broo}{\tauBo}$. All numerical results are obtained with this modification, but for the sake of clarity the lifetimes do not appear explicitly in the formulae. The values used for the lifetimes are taken from~\cite{bib:PDG}.}:
\begin{eqnarray}
\label{eq:brpm}
\Brpm&=&
{1\over 2}
(\mid\Apm\mid^2+\mid\Abarpm\mid^2)
=
\mu^2{1\over 2}(\a^2+\abar^2)
\\
\label{eq:brpo}
\Brpo&=&\mu^2
\\
\label{eq:boo}
\Broo&=&{1\over 2}
(\mid\Aoo\mid^2+\mid\Abaroo\mid^2)
=
\mu^2{1\over 2}
\left( 2+{1\over 2}(\a^2+\abar^2)-\sqrt{2}(\a\cc+\abar\ccbar) \right)
\\
\label{eq:cpm}
\Cpm&=&
{\mid\Apm\mid^2-\mid\Abarpm\mid^2
\over
\mid\Apm\mid^2+\mid\Abarpm\mid^2}
=
{\a^2-\abar^2
\over
\a^2+\abar^2}
\\
\label{eq:spm}
\Spm&=&
{-2{\rm Im}(\Apm{\Abarpm}^\ast)\over\mid\Apm\mid^2+\mid\Abarpm\mid^2}
={2\a\abar\over\a^2+\abar^2}\sin(2\alphaeff)
=\sqrt{1-\Cpm^2}\sin(2\alphaeff)
\\
\label{eq:coo}
\Coo&=&{\mid\Aoo\mid^2-\mid\Abaroo\mid^2\over\mid\Aoo\mid^2+\mid\Abaroo\mid^2}
=
{
  {1\over 2}(\a^2-\abar^2)-\sqrt{2}(\a\cc-\abar\ccbar)
\over
2+{1\over 2}(\a^2+\abar^2)-\sqrt{2}(\a\cc+\abar\ccbar)
}
\end{eqnarray}
where it proves convenient to use the two quantities $\cc$ and $\ccbar$ defined as:
\begin{eqnarray}
\cc&=&\cos(\phi),~~~~\phi~=~\alpha_0-\delta \\
\ccbar&=&\cos(\phibar),~~~~\phibar~=~\alpha_0+\delta-2\alphaeff ~.
\end{eqnarray}
The quantity $\Soo$, the analogous for $\pi^0\pi^0$ of~$\Spm$, could also be considered. Its expression is :
\begin{equation}
\Soo={
2\sin(2\alpha_0)+{\a\abar}\sin(2\alphaeff)
-{\a\sqrt{2}}\sin(\alpha_0+\delta)
-{\abar\sqrt{2}}\sin(\alpha_0-\delta+2\alphaeff)
\over
2+{1\over 2}(\a^2+\abar^2)-\sqrt{2}(\a\cc+\abar\ccbar)
} ~.
\end{equation} 
However, its measurement (through Dalitz decays of at least one of the $\piz$) implies huge statistics and it is assumed that $\Soo$ is not measurable. The six remaining measurements 
$\Brpm$, $\Brpo$, $\Broo$, $\Spm$, $\Cpm$ and $\Coo$ can in principle fix the six unknowns, including~$\alpha$, at least up to multiple solutions.
The phases $\alphaeff$, $\delta$ and~$\alpha_0$ enter the above six Eqs.(\ref{eq:brpm}-\ref{eq:coo}) only through $\cc$, $\ccbar$ and $\sin(2\alphaeff)$,
which are invariant through the three changes
$\phi\leftrightarrow-\phi$,
$\phibar\leftrightarrow-\phibar$
and 
$2\alphaeff\leftrightarrow\pi-2\alphaeff$.
Hence,
there are 8 mirror solutions in the range $[0,\pi]$
(16 in the full range $[0,2\pi]$).
These 8 mirror solutions are explicited in Table~\ref{tab:mirrors} (up to a $\pm\pi$ shift to bring them back in the range $[0,\pi]$)
where the first row indicates the true solution for the triplet ($\alpha$,$\delta$,$\alphaeff$).

\begin{table}[ht]
\begin{center}
\begin{tabular}{|c|c|c|c|c|c||}
\hline
Solutions   & $\alpha                  $ & $\delta                  $ & $\alphaeff  $ & $ \phi$ & $ \phibar$ \\ \hline\hline   
1          & $\alpha_0                $ & $\delta                  $ & $\alphaeff$ & $ \phi$ & $ \phibar$  \\ \hline      
2          & $\delta                  $ & $\alpha_0                $ & $\alphaeff$ & $-\phi$ & $ \phibar$  \\ \hline
3          & $-\alpha_0+2\alphaeff$ & $-\delta+2\alphaeff  $ & $\alphaeff$ & $-\phi$ & $-\phibar$  \\ \hline  
4          & $-\delta+2\alphaeff  $ & $-\alpha_0+2\alphaeff$ & $\alphaeff$ & $ \phi$ & $-\phibar$   \\ \hline\hline  
5          & $\pio-\alpha_0           $ & $\pio-\delta             $ & $\pio-\alphaeff$ & $-\phi$ & $-\phibar$ \\ \hline
6          & $\pio-\delta             $ & $\pio-\alpha_0           $ & $\pio - \alphaeff$ & $ \phi$ & $-\phibar$  \\ \hline  
7          & $\pio+\alpha_0-2\alphaeff$ & $\pio+\delta-2\alphaeff$ & $\pio - \alphaeff$ & $ \phi$ & $ \phibar$  \\ \hline
8          & $\pio+\delta-2\alphaeff$ & $\pio+\alpha_0-2\alphaeff$ & $\pio-\alphaeff$ & $-\phi$ & $ \phibar$     \\ \hline     
\hline
\end{tabular}
\end{center}
\caption{\it Mirror solutions for $\alpha$ in the range $[0,\pi]$. All the 8 triplets ($\alpha,\delta,\alphaeff$) yield the same values for $\cc$, $\ccbar$ and $\sin(2\alphaeff)$. Solutions (5) to (8) are just $\pi/2$ minus solutions (1) to (4). The 8 mirror solutions are strictly equivallent. They cannot be distinguished if no input is added to the ones of Eqs.(\ref{eq:brpm}-\ref{eq:coo}), like $\Soo$ for example.}
\label{tab:mirrors}
\end{table}
It is worth remarking that the mirror solutions appear in pairs symmetrical around~$\alphaeff$: 
\begin{eqnarray}
\label{eq:pair1}
\alpha(1)-\alphaeff&=&\alphaeff-\alpha(3)
\\
\label{eq:pair2}
\alpha(2)-\alphaeff&=&\alphaeff-\alpha(4)
\end{eqnarray}
and that, if the mirror solutions closest to the true value $\alpha_0$ are solutions (2)-(3)-(4), denoting them $\alpha(ii,iii,iv)$ according to their closeness rank:
\begin{equation}
\mid\alpha(ii)-\alpha_0\mid\le\mid\alpha(iii)-\alpha_0\mid\le\mid\alpha(iv)-\alpha_0\mid ~.
\end{equation}
It follows, from Table~\ref{tab:mirrors}, that the fourth-rank mirror solution cannot be too close, nor too far away from $\alpha_0\equiv\alpha(i)$, since it satisfies the double inequality:
\begin{equation}
\label{eq:badnews}
2 \mid\alpha(ii)-\alpha_0\mid\le\mid\alpha(iv)-\alpha_0\mid\le 2\mid\alpha(iii)-\alpha_0\mid ~.
\end{equation}
Therefore, if the second-rank mirror solution $\alpha(ii)$ is near $\alpha_0$, while the third-rank mirror solution $\alpha(iii)$ is also creeping close, then the fourth-rank solution $\alpha(iv)$ is nor far either: in that case, $\alpha_0$ will not be precisely measured. This topic is discussed in Section~\ref{sec:GrossmanQuinn} in the light of the Grossman-Quinn bound.

\section{Algebra}
\label{sec:algebrasection}
In this section one deals with some stark mathematics. The algebra involved being quite simple, only the results are presented: they are important for the next sections. It is demonstrated in passing that the sign of~$\cos(2\alphaeff)$ is firmly established by \babar\ data, and the current CKM fit results.
Below, are discussed the two types of bounds one may find in the literature: lower (and upper) bounds on $\Broo$, and upper bounds on $\mid\alpha-\alphaeff\mid$. These bounds are of some interest, but they can also be misleading. Furthermore, because of the limited precision on the measurements they use, all bounds (and the new bound of Eq.(\ref{eq:rebound}) is no exception) do not provide more than qualitative insights. To turn a bound value into a quantitative statement, one should append the CL it refers to, and this implies to have recourse to fits, which is done in Section~\ref{sec:perspective}.

\subsection{New bounds on $\Broo$}
\label{sec:newbounds}
New bounds on $\Broo$ are presented, using the current knowledge on~$\alpha$. The value determined by the standard CKM fit is $96 \pm 13^o$ ($1.67 \pm 0.23~\rad$)~\cite{bib:ckmfitterweb}. Using $\alpha$ presents several interests:
\begin{itemize}
\item{} to help cornering $\Broo$. 
For the time being, the question is not yet to pinpoint~$\alpha$ thanks to $B\rightarrow\pi\pi$ decays,
but to evaluate the chances one has to do so in the close or distant future.
With that respect, one is longing for a robust estimate of~$\Broo$,
and every tools at hand should be used to secure it.
The information on~$\alpha$ must be used to this end as well,
\item{} to derive the sign of~$\cos(2\alphaeff)$ (Section~\ref{sec:toughBoundsection}),
\item{} to understand the behavior of the mirror solutions (Section~\ref{sec:scenariospace}).    
\end{itemize}
Assuming that~$\alpha$ is known, the set of equations for the $B\rightarrow\pi\pi$ amplitudes implies the following lower bounds for the $B^0\rightarrow\pi^0\pi^0$ branching ratio~\footnote{They are obtained by deriving $\Broo$ with respect to $\delta$.}:
\begin{eqnarray}
\label{eq:newbound}
\newBoundl&=&\Brpo+{1\over 2}\Brpm - \sqrt{\Brpo\Brpm(1+O_{\pm})} 
\\
\label{eq:newboundbis}
O_{\pm}&=&\sqrt{1-\Cpm^2}\ \cos(2(\alpha_0-\alphaeff)) 
\\
&=&\pm\sqrt{(1-\sin^2(2\alpha_0))(1-\Cpm^2-\Spm^2)}+\sin(2\alpha_0)\Spm
\end{eqnarray}
where $\pm$ in $O_{\pm}$ is the sign of the product $\cos(2\alphaeff)\cos(2\alpha_0)$.

They come with the upper bounds $\newBoundu$ obtained by reversing the sign in front of the square root. However, the upper bounds are not very relevant, being much weaker than the direct experimental upper limit.

\subsubsection{The reBound}
\label{eq:reBoundsection}
The bound $\reBoundl$ obtained when $\cos(2\alphaeff)\cos(2\alpha_0)$ is taken to be positive is referred to as the reBound: 

\begin{center}
{\fbox {{\parbox{0.7\textwidth}{
\begin{equation}
\label{eq:rebound}
\reBoundl = \Brpo+{1\over 2}\Brpm-\sqrt{\Brpo\Brpm(1+O_{+})}
\end{equation}
\begin{equation}
O_{+} = \sqrt{(1-\sin^2(2\alpha_0))(1-\Cpm^2-\Spm^2)}+\sin(2\alpha_0)\Spm
\end{equation}
}}}\par }
\end{center}
The alternative bound $\toughBoundl$ is considered in the next section.

The reBound can be compared with previous bounds~\cite{bib:GQ,bib:Jerome}, the strongest lower one being the Gronau-London-Sinha-Sinha (GLSS) bound~\cite{bib:GLSS}:
\begin{equation}
\label{eq:bound}
\BrooGLSSl=\Brpo+{1\over 2}\Brpm-\sqrt{\Brpo\Brpm \left(1+\sqrt{1-\Cpm^2}\right)}
\end{equation}
which minimal value with respect to $\Cpm$ is attained for $\Cpm=0$:
\begin{equation}
\label{eq:boundmin}
\BrooGLSSl[\rm min]=\Brpo\left(1-\sqrt{\Brpm\over 2\Brpo}\right)^2 ~.
\end{equation}
It comes in pair with the upper bound $\BrooGLSSu$ obtained by reverting the sign in front of the square root. However, as obvious from Eq.(\ref{eq:newboundbis}) the  reBound is more stringent than the GLSS bound, which it reproduces if no knowledge on~$\alpha$ is assumed.
The largest difference is obtained with the \sBelle\ measu\-re\-ments
($\Cpm^2+\Spm^2=1$), which, taken together with the value $\alpha_0~=~90^o$, imply the new lower limit $\Broo>2.3$, while the weaker bound of Eq.(\ref{eq:bound}) is $\Broo>0.7$ (see Table~\ref{tab:bounds}).

Using unadorned standard error propagation to estimate the precision of the bound, one gets for the reBound, $\alpha~=~96^o$, and,

\noindent \babar\ values: 
\begin{equation}
\label{eq:bounderrorBa}
\reBoundl=0.6^{+0.5}_{-0.3}[\Brpo]^{+0.2}_{-0.1}[\Brpm]^{+0.1}_{-0.0}[\Spm]^{+0.1}_{-0.0}[\Cpm]^{+0.2}_{-0.0}[\alpha]
\end{equation}
\sBelle\ values:
\begin{equation}
\label{eq:bounderrorBe}
\reBoundl=1.9^{+0.7}_{-0.5}[\Brpo]^{+0.1}_{-0.0}[\Brpm]^{+0.1}_{-0.1}[\Spm]^{+0.0}_{-0.0}[\Cpm]^{+0.6}_{-0.4}[\alpha]
\end{equation}
and world average values:
\begin{equation}
\label{eq:bounderrorWA}
\reBoundl=1.1^{+0.3}_{-0.3}[\Brpo]^{+0.1}_{-0.1}[\Brpm]^{+0.0}_{-0.2}[\Spm]^{+0.2}_{-0.1}[\Cpm]^{+0.4}_{-0.2}[\alpha]
\end{equation}
where the source of the uncertainty on the bound is specified within brackets. Whereas this tells that the main negative inaccuracy on the reBound value stems from the branching ratio $\Brpo$,
one cannot infer much else: the errors are large and correlated,
hence they cannot be added in quadrature.
Anyway, 
one should keep in mind that
the actual 90\% CL lower limit on $\Broo$ can hardly be guessed from back of the envelope calculations.

The numerical results for the reBound for \babar, Belle and world average are summarized in Table~\ref{tab:bounds} for three different $\alpha_0$ values. In this table are also given numerical values of other bounds presented in the following sections.
\begin{table}[ht]
\begin{center}
\begin{tabular}{|c||c||c|c|c||c|c|c||c||c||}
\hline
\multicolumn{1}{|c||}{Bounds} &
\multicolumn{1}{|c||}{$\BrooGLSSl$} &
\multicolumn{3}{|c||}{$\reBoundl$: reBound} &
\multicolumn{3}{|c||}{$\toughBoundl$: tough Bound} &
\multicolumn{1}{|c||}{$\Broo/\Brpo$} &
\multicolumn{1}{|c||}{$\sinmax$} \\
\hline
$\alpha$ & & $90^o$& $96^o$& $110^o$& $90^o$& $96^o$& $110^o$ & & \\ \hline
\babar\      &$0.6$ & $0.7$& $0.6$ & $0.6$& $5.9$& $5.2$ & $3.8$& $0.4$ & $0.4$\\ \hline
\sBelle\      &$0.7$ &$2.3$& $1.9$ & $1.2$& $2.3$& $1.9$ & $1.2$& $0.4$ & $0.3$\\ \hline
W.A.       &$0.7$ & $1.3$& $1.1$ & $0.7$& $3.8$ & $3.3$ & $2.2$& $0.4$ & $0.4$\\ \hline\hline 
\end{tabular}
\end{center}
\caption{\it Numerical values for the different bounds: the GLSS bound, the reBound, the tough Bound, the original (approximate) Grossman-Quinn bound and the exact Grossman-Quinn bound. Branching ratios are quoted in the $10^{-6}$ unit. The last two values are discussed in Section~\ref{sec:GrossmanQuinn}.}
\label{tab:bounds}
\end{table}

\subsubsection{The tough Bound}
\label{sec:toughBoundsection}
The hypothesis that the sign of~$\cos(2\alphaeff)$ is opposite to the sign of~$\cos(2\alpha_0)$ is now made. A more stringent bound is then obtained:
\begin{equation}
\label{eq:toughBound}
\toughBoundl=\Brpo+{1\over 2}\Brpm-\sqrt{\Brpo\Brpm(1+O_-)} 
\end{equation}
which is referred to as the tough Bound. For \sBelle, since $\Cpm^2+\Spm^2=1$, the tough Bound is identical to the reBound. However,
for \babar\ and for the world average, reverting the sign makes a huge difference. The precision on the tough Bound for $\alpha~=~96^o$ is, for the \babar\ values:
\begin{equation}
\label{eq:toughbounderrorBaBar}
\toughBoundl=
 5.2^{+1.0}_{-0.9}[\Brpo]^{+0.2}_{-0.2}[\Brpm]^{+0.5}_{-0.5}[\Spm]^{+0.1}_{-0.2}[\Cpm]^{+0.9}_{-0.9}[\alpha]
\end{equation}
for \sBelle:
\begin{equation}
\label{eq:toughbounderrorBe}
\toughBoundl=
 1.9^{+0.7}_{-0.5}[\Brpo]^{+0.1}_{-0.0}[\Brpm]^{+0.1}_{-0.1}[\Spm]^{+0.0}_{-0.0}[\Cpm]^{+0.6}_{-0.4}[\alpha]
\end{equation}
and for the world average:
\begin{equation}
\label{eq:toughbounderrorwa}
\toughBoundl=
 3.3^{+0.5}_{-0.4}[\Brpo]^{+0.0}_{-0.0}[\Brpm]^{+0.4}_{-0.0}[\Spm]^{+0.2}_{-0.3}[\Cpm]^{+0.7}_{-0.7}[\alpha] ~.
\end{equation}
These tough Bound values are, for \babar\ and W.A., firm enough to be barely consistent with the measurement of~$\Broo$ (Table~\ref{tab:inputs}): the hypothesis $\cos(2\alphaeff)\cos(2\alpha_0)<0$ is disfavored. On the other hand, the sign of~$\cos(2\alpha)$ is well defined to be negative by the current state of the art of CKM fits. Hence, the tough Bound suggests $\cos(2\alphaeff)<0$.

\subsection{The $\Coo$ function}
\label{sec:coo}
The asymmetry term $\Coo$ (Eq.(\ref{eq:coo})) has not been yet measured due to the insufficient luminosity. From the knowledge of~$\alpha$, once the three branching ratios $\Brpm$, $\Brpo$, $\Broo$ and the two asymmetries $\Spm$ and $\Cpm$ are measured, $\Coo$ can be determined, up to a fourfold ambiguity:

\begin{center}
{\fbox {{\parbox{0.9\textwidth}{
\begin{eqnarray}
\label{eq:master}
\Coo^{\pm;+}&=&
\frac{1}{\Broo(1+O_{+})} \Bigg\{
-\Cpm \left(\Brpo-{\Brpm\over 2}O_{+}-\Broo\right)\nonumber \\
&\phantom{=}&\pm
\sqrt{(\Broo-\reBoundlepscos)(\reBounduepscos-\Broo)(1-O_{+}^2-\Cpm^2)} \Bigg\}
\end{eqnarray}
}}}}
\end{center}
One obtains the two other solutions $\Coo^{\pm;-}$ by using $O_{-}$ instead of $O_{+}$ in the above expression.
For $\alpha_0~=~90^o$ some simplifications occur:
\begin{eqnarray}
\label{eq:masterpio}
O_{\pm}&=&\pm\sqrt{1-\Cpm^2-\Spm^2} \\
\label{eq:coo90}
\Coo^{\pm;\pm} &=&
\frac{1}{\Broo(1+O_{\pm})} \Bigg\{-\Cpm\left(\Brpo-{\Brpm\over 2}O_{\pm}-\Broo\right)
\nonumber \\
&\phantom{=}&
\pm\Spm\sqrt{(\Broo-\reBoundlepscos)(\reBounduepscos-\Broo)}\Bigg\} ~.
\end{eqnarray}
This multivalued function is drawn on Fig.~\ref{fig:cooboo}, for \babar, \sBelle\ and the world average, with $\alpha~=~90$, $96$ and $110^o$. The $\Broo$ axis spans the range up to $15 \times 10^{-6}$ despite the known experimental upper bound to allow a wide view on the $\Coo(\Broo)$ function. Each of the two closed loops corresponds to a given choice for the sign of $\cos(2\alphaeff)\cos(2\alpha_0)$: the plain lines correspond to $O_{+}$, the dotted ones to $O_{-}$. The upper and lower parts of each closed loop correspond to the $\pm$ sign of Eq.(\ref{eq:master}). The vertical lines indicate the upper and lower bounds of the allowed $\Broo$ range. The outermost vertical lines correspond to the reBound lower (and upper) limit, while the inner lines correspond to the tough Bound limits.

For \sBelle, $\Cpm^2+\Spm^2$ being taken equal to unity and then the reBound being equal to the tough Bound, the two closed loops are superimposed.

For \babar\ and the world average, both the closed loops and their upper and lower components are widely split apart. For \babar\ and the W.A., the inner part open when~$\alpha$ becomes larger, it is the contrary for the outer curves: the larger~$\alpha$ becomes, the closer they get from each other.

The variation of the value of~$\alpha$ affects the $\Coo(\Broo)$ function domain for \babar, Belle and W.A. In the experimentally allowed region ($\Broo < 4 \times 10^{-6}$), both negative and positive values are possible for $\Coo$. Here again, such indications are to be taken with a grain of salt since the uncertainties on the measurements are not accounted for in the discussion (but they are in Section~\ref{sec:StandarModelstestsection}).

\begin{figure}[h!]
\begin{center}
\mbox{\epsfig{file=tune_babar.eps_alpha90,width=0.33\linewidth}
\epsfig{file=tune_belle.eps_alpha90,width=0.33\linewidth}
\epsfig{file=tune_wa.eps_alpha90,width=0.33\linewidth}}
\mbox{\epsfig{file=tune_babar.eps_alpha96,width=0.33\linewidth}
\epsfig{file=tune_belle.eps_alpha96,width=0.33\linewidth}
\epsfig{file=tune_wa.eps_alpha96,width=0.33\linewidth}}
\mbox{\epsfig{file=tune_babar.eps_alpha110,width=0.33\linewidth}
\epsfig{file=tune_belle.eps_alpha110,width=0.33\linewidth}
\epsfig{file=tune_wa.eps_alpha110,width=0.33\linewidth}}
\end{center}
\caption{\it The multivalued function $\Coo^{\pm;\pm}=f(\Brpm,\Brpo,\Broo,\Spm,\Cpm,\alpha)$ is drawn as a function of~$\Broo$ using $\alpha~=~90$ (top), $96$ (middle) as measured by the standard CKM fit, and $110^o$ (bottom), for the measurements from \babar\ (left), \sBelle\ (middle), and the world average (right). The outer limits correspond to $\cos(2\alphaeff)\cos(2\alpha_0)>0$ (i.e. the reBound of Eq.(\ref{eq:rebound})) and the inner limits to $\cos(2\alphaeff)\cos(2\alpha_0)<0$ (i.e. the tough Bound of Eq.(\ref{eq:toughBound})).}
\label{fig:cooboo}
\end{figure}

\subsection{Explicit expression for~$\alpha$}
\label{sec:closeformforalpha}
The identity $2\alpha~=~2\alphaeff+\phi+\phibar$ leads to the following expression for~$\alpha$:
\begin{equation}
\label{eq:tanalpha}
\tan{\alpha}=
{
\sin(2\alphaeff)\ccbar+\cos(2\alphaeff)\ssbar+\ss
\over
\cos(2\alphaeff)\ccbar-\sin(2\alphaeff)\ssbar+\cc
}
\end{equation}
where all quantities on the right hand side can be expressed in term of the observables as follow:
\begin{eqnarray}
\sin(2\alphaeff)&=&{\Spm\over\sqrt{1-\Cpm^2}}
\\
\cos(2\alphaeff)&=&\pm \sqrt{1-\sin^2(2\alphaeff)}
\\
\cc&=&\cos(\phi)={\Brpo+\Brpm(1+\Cpm)/2-\Broo(1+\Coo)
\over
\sqrt{2\Brpm\Brpo(1+\Cpm)}}
\\
\ss&=&\sin(\phi)=\pm\sqrt{1-\cc^2}
\\
\ccbar&=&\cos(\phibar)={\Brpo+\Brpm(1-\Cpm)/2-\Broo(1-\Coo)
\over
\sqrt{2\Brpm\Brpo(1-\Cpm)}}
\\
\ssbar&=&\sin(\phibar)=\pm\sqrt{1-\ccbar^2}
\end{eqnarray}
where the eightfold ambiguity in the range $[0,\pi]$ is explicited by the three arbitrary signs.
One recovers the GLSS bound by requesting that both $\cc$ and $\ccbar$ are smaller than one.
At the edge of the allowed $\Broo$ range,
for $\Broo=\BrooGLSSl$, 
one gets:
\begin{eqnarray}
\phi&=&\phibar=0
\rightarrow\alpha=\alphaeff=\delta
\\
\Coo&=&{1\over \BrooGLSSl}\left({\Brpm\over 2}\Cpm+\sqrt{\Brpo\Brpm\left(1-\sqrt{1-\Cpm^2}\right)}\right)
\\
\reBoundl&=&\BrooGLSSl ~.
\end{eqnarray}
Being in an Universe where $\Broo$ sits right at the GLSS bound may be viewed,
at first glance, as good news. Indeed, in that case one has the nice equality $\alpha~=~\alphaeff$, as if there were no Penguins, and, as an odd bonus, one still savors direct CP violation since neither $\Cpm$ nor $\Coo$ has to vanish.
However, to be in an Universe where $\Broo$ is simply close to the GLSS bound is bad news!
Indeed, in that case, 
in the neighbourhood of $\alpha\simeq\alphaeff$ lurk the three mirror solutions (2)-(3)-(4), while the other four solutions (5)-(6)-(7)-(8) are thrown apart.
Furthermore, by construction, when one is close to the GLSS bound $\partial\Broo/\partial\delta=0$, therefore the locations of the mirror solutions (2) and (4) vary wildly with mild changes of $\Broo$. This subject is discussed in details in Section~\ref{sec:scenariospace} and illustrated on Fig.~\ref{fig:switchyard}. The implication that $\Broo=\BrooGLSSl$ forces $\alpha~=~\alphaeff$ is reminiscent of the Grossman-Quinn bound, which is discussed in the next section.

\subsection{The Grossman-Quinn bound revisited}
\label{sec:GrossmanQuinn}
%
The Grossman-Quinn (GQ) bound was introduced assuming that no knowledge of $\alpha$ is available and $\Coo$ is not measured. It is mostly known (and used) as:
\begin{equation}
\label{eq:GQbound}
\sin^2(\alpha-\alphaeff)\le {\Broo\over\Brpo} ~.
\end{equation}
This bound appears quite peculiar owing to the results of the previous section where it was shown that $\alpha$ equals $\alphaeff$ if $\Broo$ reaches the GLSS bound, not if $\Broo$ is small.
In effect, the GLSS bound is not necessarily a small number:
one may thus be able to set a tight limit on the difference $\mid\alpha-\alphaeff\mid$ even though $\Broo$ is not small. 
Mathematically,
the constraint on $\mid\alpha-\alphaeff\mid$
does not stem from the size of the branching ratio,
but from the proximity of the latter to its lower (or upper) bound,
if no knowledge is assumed on~$\alpha$.
Physically, there is no reason to expect $\alpha\simeq\alphaeff$ except if Penguin diagrams turn out to be very small, which is not a strong hope. If it is not the case then only an unfathomable numerical accident may lead to $\Broo\simeq\BrooGLSSl$. In fact, a straight maximization of $\sin^2(2(\alpha-\alphaeff))=\sin^2(\phi+\phibar)$ with respect to $\Coo$ yields:

\begin{equation}
\Coo[{\rm max}]=-\Cpm\left({\Brpm\over 2\Broo}\right)
\left({
\Brpo-{\Brpm\over 2}+\Broo
\over
\Brpo+{\Brpm\over 2}-\Broo
}\right)
\end{equation}

\begin{center}
{\fbox {{\parbox{0.8\textwidth}{
\begin{equation}
\label{eq:sinmax}
\sin^2(\alpha-\alphaeff)
~\le~
\sinmax
=
{
\left(
\Broo-\BrooGLSSl
\right)
\left(
\BrooGLSSu-\Broo
\right)
\over
2\Brpm\Brpo\sqrt{1-\Cpm^2}
}
\end{equation}
}}}}
\end{center}
This is not a new result:
although the end product appears in a quite different form,
it is derived in the same way in Ref.\cite{bib:GLSS},
and is approached in Ref.\cite{bib:Jerome}. 
It is also present,
albeit on a low key and in a very different looking form,
in the original paper~\cite{bib:GQ} itself~\footnote{Except that, in the second parenthesis, $\Broo$ is replaced by its lower bound $\BrooGLSSl$,
which underestimates the strength of the bound on $\mid\alpha-\alphaeff\mid$.}.  
To recover the GQ bound Eq.(\ref{eq:GQbound}) from Eq.(\ref{eq:sinmax}), one may write:
\begin{equation}
\label{eq:GQsinmax}
\sinmax={\Broo\over\Brpo}
+\left(
{1\over\sqrt{1-\Cpm^2}}
-1
\right)
\left(
{\Broo\over\Brpo}-{1\over 2}
\right)
-{
\left(\Broo+{\Brpm\over 2}-\Brpo\right)^2
\over
2\Brpo\Brpm\sqrt{1-\Cpm^2}
} ~.
\end{equation}
With \babar, Belle and W.A. values both (competing) additionnal terms are small by chance. The second term is very small for two reasons: because (for \babar) $\Cpm^2\simeq 0.04$, and because (for all) $\Broo/\Brpo \simeq 0.4 \simeq 1/2$. The third term is small because $\Broo$ is closed to $\Brpo-\Brpm/2$.

Hence the validity of the approximation of Eq.(\ref{eq:GQbound}), when it applies, also deprives the GQ bound of its interest: on the one hand, it yields a weak bound, and on the other hand, the bound is irrelevant since $\Broo$ and $\Coo$ are likely to be measurable, making the full isospin analysis tractable.

Notwithstanding the above considerations,
to use Eq.(\ref{eq:GQbound}) entails an error which can go in either directions,
as can be seen from Eq.(\ref{eq:GQsinmax}).
It can lead to overestimate\footnote{
The Charles bound~\cite{bib:Jerome} 
and the second GQ bound~\cite{bib:GQ}
are intermediate results between the GQ
and the GLSS bounds which can only understimate the strenght of the bound on $\mid\alpha-\alphaeff\mid$.
} 
or to underestimate the bound on $\mid\alpha-\alphaeff\mid$.
Although with the present data, the effect on the calculated values (Table~\ref{tab:bounds}) is, by chance, not a large one, there is no reason not to use the $\sinmax$ bound: the underlying assumptions are identical. 

Before concluding this section,
it is appropriate to remind that setting a bound on $\mid\alpha-\alphaeff\mid$ is
not an alternative to the full isospin analysis:
the latter encompasses the first.
It is also worth repeating that a tight bound on $\mid\alpha-\alphaeff\mid$ also implies
that three mirror solutions are zeroing on~$\alpha$: 
it is good news if they all get much closer than the statistical power of the isospin analysis,
it is very bad news if they just stay at bay within one or two sigma's of~$\alpha$. In that case, the measurement is spoilt.

Remark: Finally, one should keep in mind that Eq.(\ref{eq:sinmax}) does not refer to $\alpha_0$ but to any of the eight solutions for~$\alpha$. In particular, since $\alpha_0$ and $\delta$ are degenerate in the isospin analysis, $\delta$ obeys the constraint of Eq.(\ref{eq:sinmax}) as well. More precisely, if one considers for definiteness the first four mirror solutions for~$\alpha$, and if for example solution (2) reaches the bound $\sinmax$, then:
\begin{equation}
\sin^2(\alpha(1)-\alphaeff)=\sinmax
{1-\sqrt{1-\Cpm^2}\over 1+\sqrt{1-\Cpm^2}}
\left(
{
(\BrooGLSSu-\Broo)+(\Broo-\BrooGLSSl)
\over
(\BrooGLSSu-\Broo)-(\Broo-\BrooGLSSl)
}
\right)^2
\end{equation}
while the locations of the two other mirror solutions are fixed by Eqs.(\ref{eq:pair1}) and~(\ref{eq:pair2}). In particular, if $\Cpm$ is small, omitting the last parenthesis which is close to unity, the drift from~$\alphaeff$ of~$\alpha(1)$ is very small:
\begin{equation}
\sin^2(\alpha(1)-\alphaeff)=\sinmax\ {\Cpm^2\over 4} ~.
\end{equation} 
However, this small value is not a bound. It is only a snapshot: the value reached by~$\alpha(1)$ when $\alpha(2)$ hits the $\sinmax$ bound. If $\alpha(2)$ is not at the bound $\sin^2(\alpha(1)-\alphaeff)$ takes higher or lower values. In particular, a symmetrical situation is encountered when $\alpha(1)$ hits $\sinmax$.

\section{Perspective for the~$\alpha$ measurement}
\label{sec:perspective}
To illustrate the above relations, different scenarii are considered with possible values of $\Broo$, $\Coo$ and parameters errors in the future. In Section~\ref{sec:scenariospace} the evolution of the eight mirror solutions are studied as a function of~$\Broo$. In Section~\ref{sec:guesses}, the present world average values are used with their errors scaled down by a dreamy factor 10. That is to say a luminosity of roughly $10~\invab$ is considered both for \babar\ and Belle.
This deliberately unrealistic treatment is chosen to underline the fact that most of the difficulties to predict the outcome of the future analyses arise from the vast array of possibilities still open for the true characteristics of the $B\rightarrow\pi\pi$ decays.
They are not primarily due to the detailed performances of the analyses,
although of course,
the latter are a matter of deep concern.
In Section~\ref{sec:TheNastyScenarioSection} to illustrate this point, the \babar\ values are considered and the present analysis performances applied to a $500~\invfb$ data sample are considered for two theoretical scenarii concerning $\Broo$: a nasty one (scenario I) and a nice one (scenario II).

The results of the analysis are expressed as confidence levels on $\alpha$. A fit is performed minimizing the straight $\chi^2$
\begin{eqnarray}
\label{eq:chi2}
\chi^2&=&
\ \ \left({\Brpmexp-\Brpm\over\sigma[\Brpm]}\right)^2
+\left({\Brpoexp-\Brpo\over\sigma[\Brpo]}\right)^2
+\left({\Brooexp-\Broo\over\sigma[\Broo]}\right)^2
\nonumber \\
&\phantom{=}&
+\left({\Spmexp-\Spm\over\sigma[\Spm]}\right)^2
+\left({\Cpmexp-\Cpm\over\sigma[\Cpm]}\right)^2
+\left({\Cooexp-\Coo\over\sigma[\Coo]}\right)^2
\end{eqnarray}
to obtain Figs.~\ref{fig:alphascan},~\ref{fig:alphascan2} and~\ref{fig:alphascannasty}.

\subsection{Scenario space}
\label{sec:scenariospace}
The space of scenarii allowed by the current state of knowledge is a large one. A wide point of view is taken and six two-dimensional subsets of the parameters are embraced. These are defined by taking at face value the \babar, \sBelle\ and W.A. measurements, together with $\alpha~=~96^o$ ($1.67~\rad$).

\begin{figure}[h!]
\begin{center}
\mbox{\epsfig{file=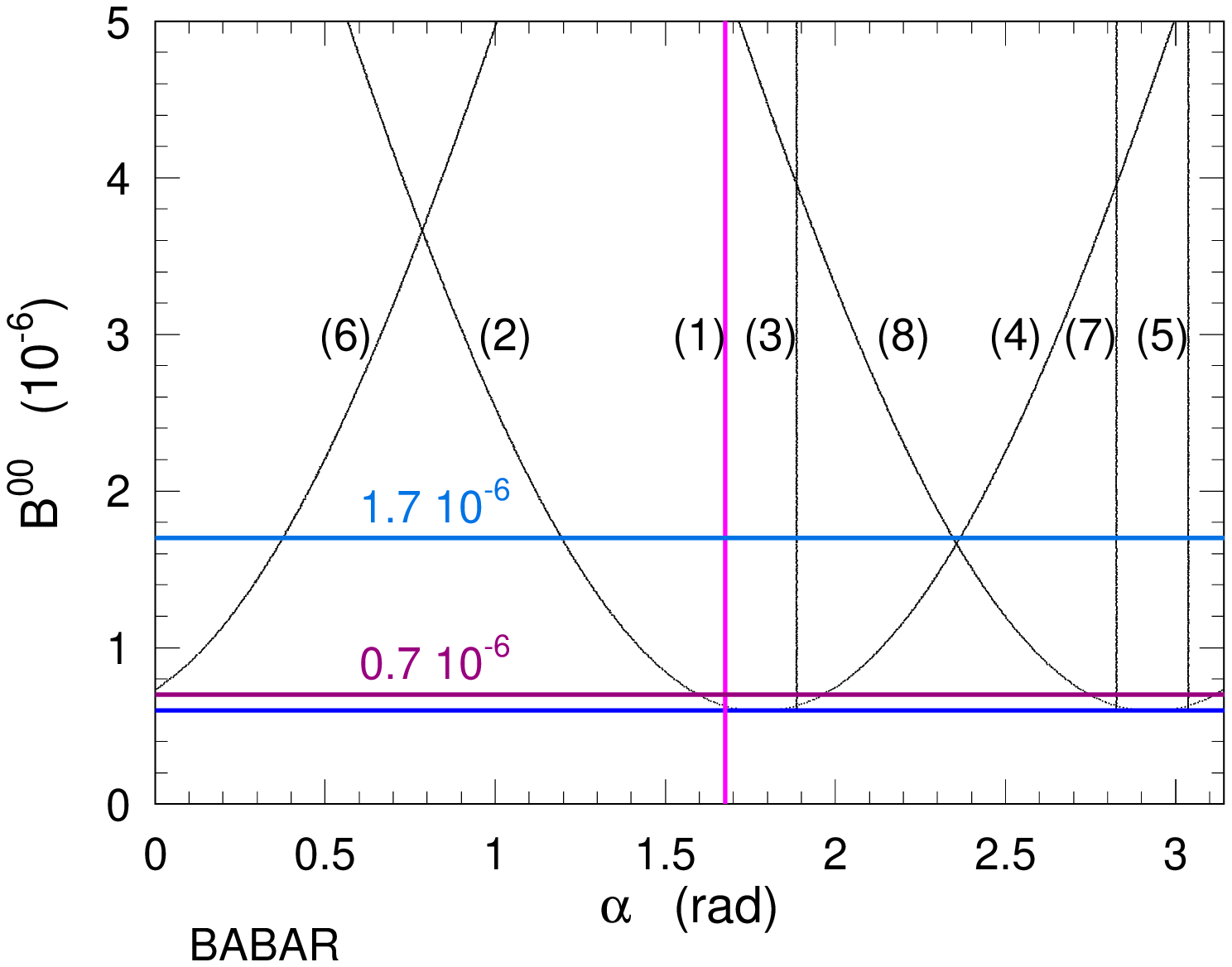,width=.4\linewidth}
\epsfig{file=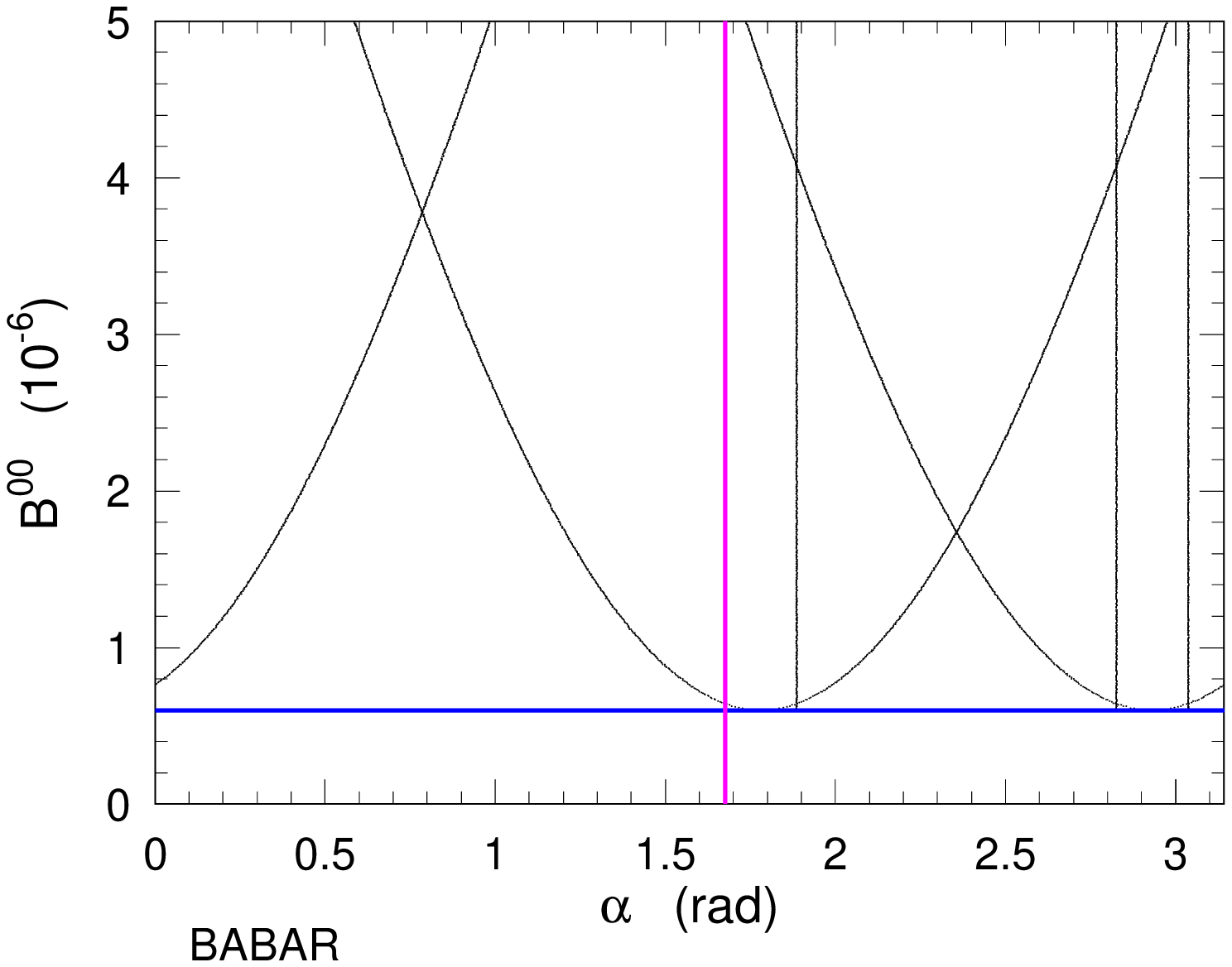,width=.4\linewidth}}
\mbox{\epsfig{file=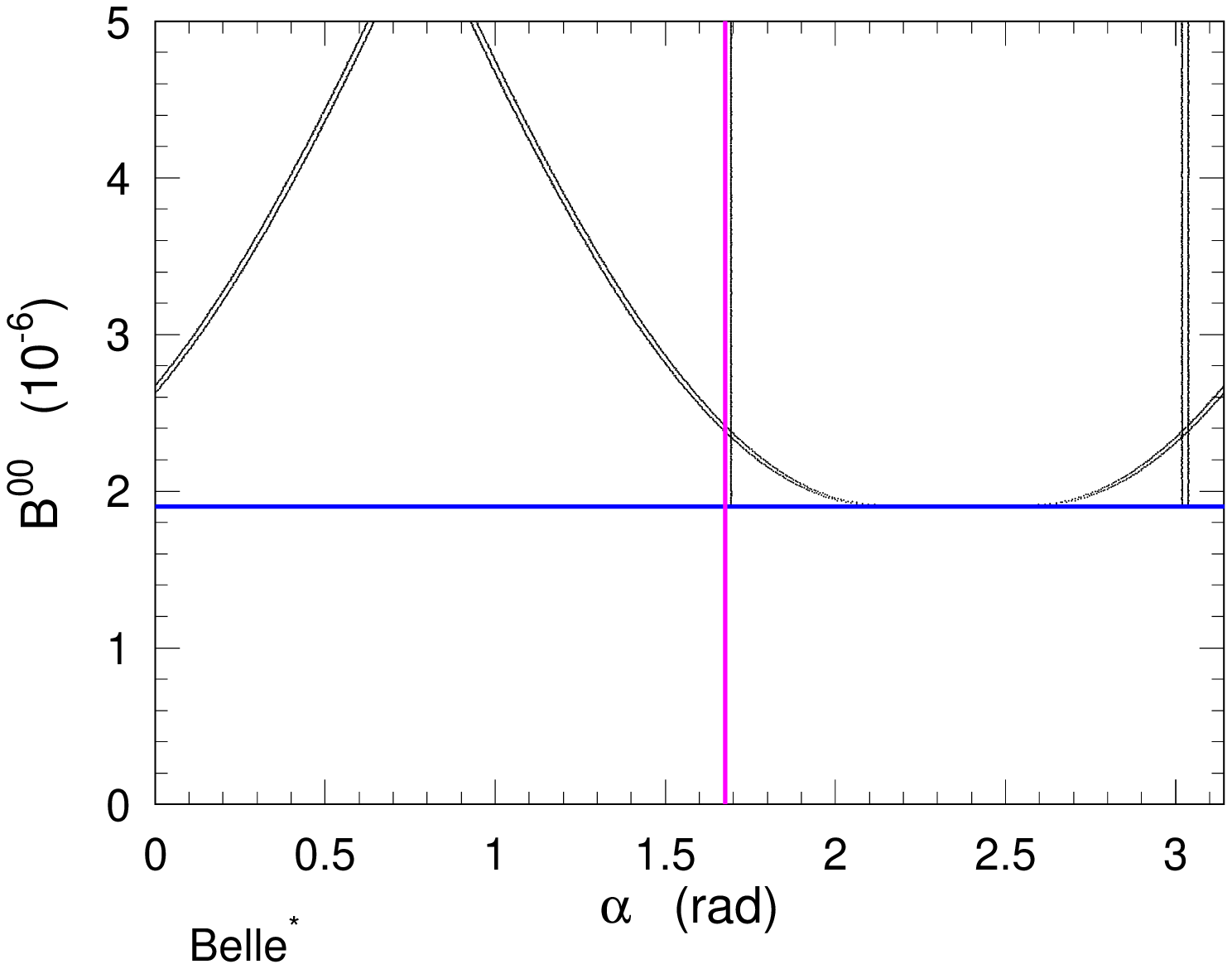,width=.4\linewidth}
\epsfig{file=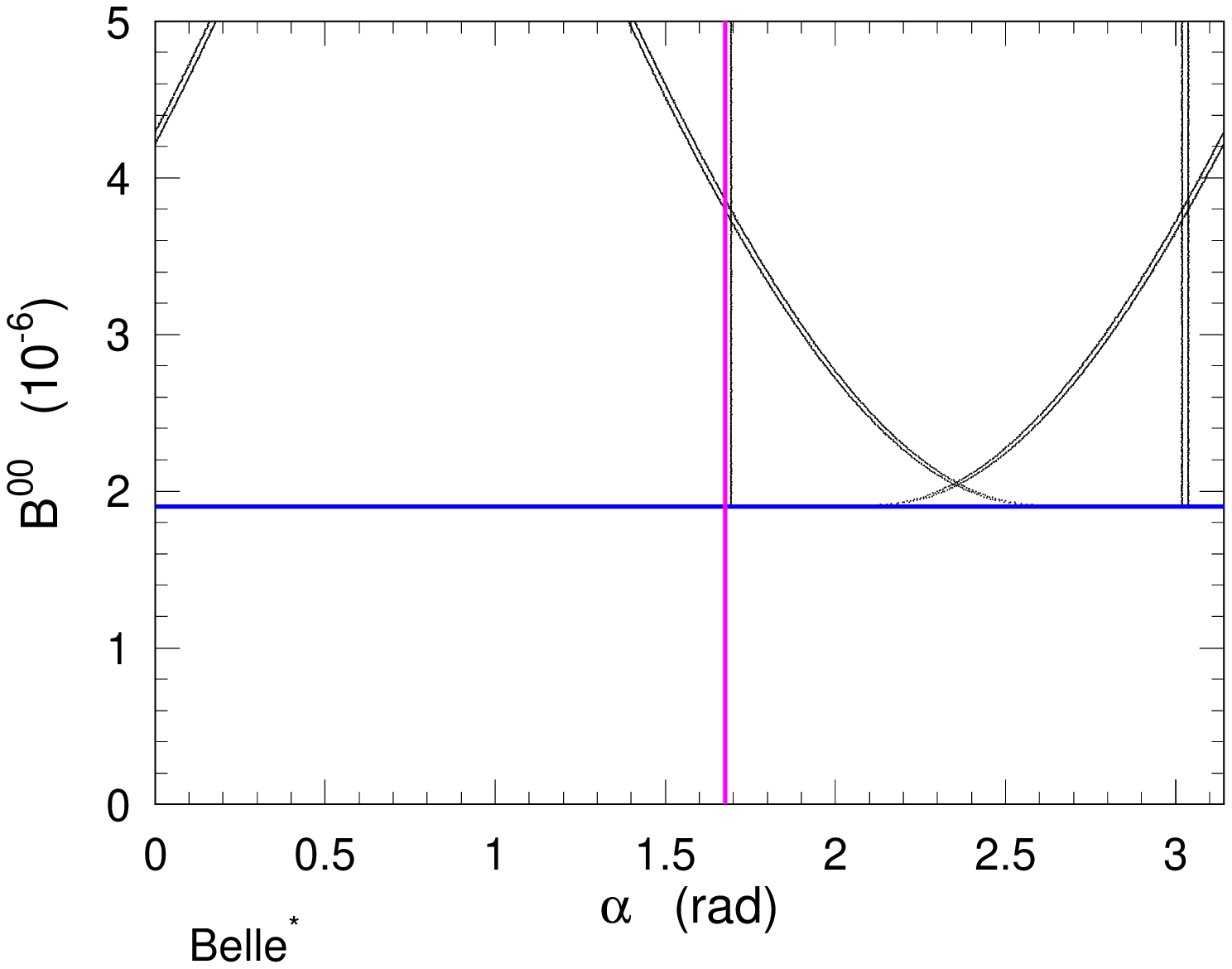,width=.4\linewidth}}
\mbox{\epsfig{file=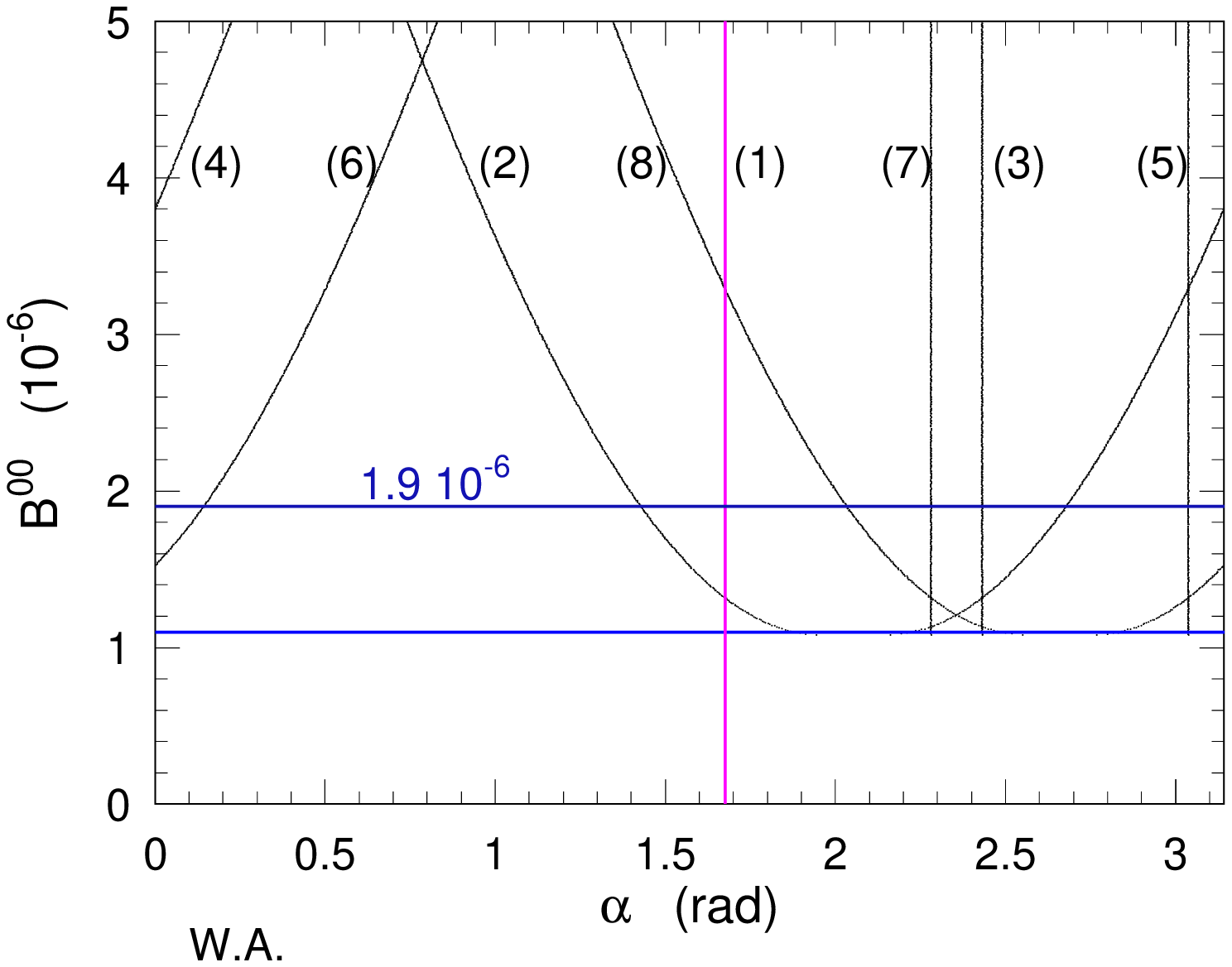,width=.4\linewidth}
\epsfig{file=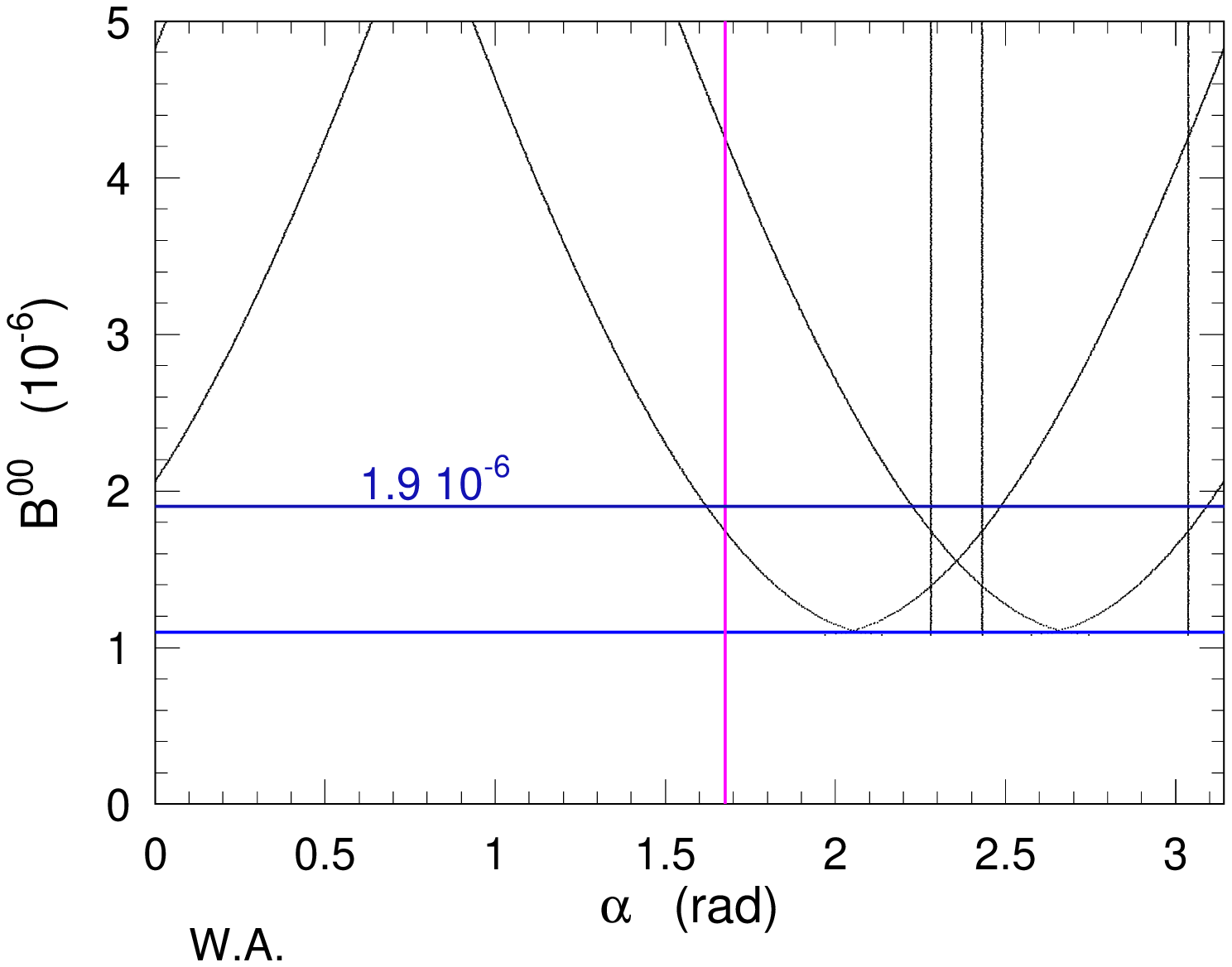,width=.4\linewidth}}
\end{center}

\vspace{-0.8cm}
\caption{\it Location of the 8 mirror solutions as a function of $\Broo$. The six sets of curves refer to \babar\ (top), \sBelle\ (middle) and the world average (bottom). Each set starts at the minimal value $\Broo=\reBoundl$. Only the solutions obtained with $\cos(2\alphaeff)$ negative (Section~\ref{sec:toughBoundsection}) are considered. The curves on the left hand side are obtained with the $\Coo$ value corresponding to the upper part of the $\Coo(\Broo)$ function (Fig.~\ref{fig:cooboo}), the ones on the right hand side with the $\Coo$ value corresponding to the lower part of the $\Coo(\Broo)$ function. The vertical lines drawn on each plot represent the true value of $\alpha$ ($96^o$) and the mirror solutions (3), (5) and (7) which are independent of $\Broo$. Four scenarii serve as illustrations in Section~\ref{sec:guesses} (W.A. using $\Broo=1.9 \times 10^{-6}$) and Section~\ref{sec:TheNastyScenarioSection} (\babar, using $\Broo=1.7$ and $0.7 \times 10^{-6}$). The corresponding horizontal lines are drawn here to indicate the locations of the 8 mirror solutions.}
\label{fig:switchyard}
\end{figure}

On Fig.~\ref{fig:switchyard} are shown the location of the 8 mirror solutions (Table~\ref{tab:mirrors}) as a function of the value assumed for $\Broo$. Each set of curves starts at the minimal value given by the reBound. For the sake of clarity, only the solutions obtained with $\cos(2\alphaeff)$ negative are considered, in agreement with the finding of Section~\ref{sec:toughBoundsection}. Otherwise, one should also consider another set of six curves, this time starting at the tough Bound.

Four of the mirror solutions given in Table~\ref{tab:mirrors} do not depend on $\Broo$: beside the true solution, they are the solutions (3), (5) and (7). For \babar\ as for \sBelle\ one of these branching ratio independent solutions is close to the true solution. Whatever $\Broo$, this nearby mirror solution has a harmful effect on the measurement. However, the current uncertainties on $\Cpm$ and $\Spm$ are large enough to allow for this mirror solution to drift away swiftly with the coming updates. For the world average, all mirror solutions are harmlessly far apart.

The location of the other four mirror solutions depend on $\Broo$, because $\Coo$ is a multivalued function of this branching ratio. The $\Coo(\Broo)$ function drawn on Fig.~\ref{fig:cooboo} shows that there exist two values of~$\Coo$ for a given $\Broo$ and for a given sign of the product $\cos(2\alphaeff)\cos(2\alpha_0)$. The three sets of curves on the left hand side are obtained with the upper corresponding values of $\Coo$ and the three sets of curves on the right hand side with the lower values of $\Coo$ (for $\cos(2\alphaeff)\cos(2\alpha_0)$ positive). The curves for the two different $\Coo$ bounce back together when $\Broo$ hits the reBound $\reBoundl$. 

The worst case scenario is when three mirror solutions gather around the true solution (ie. when $\Broo \simeq \BrooGLSSl$). With \babar\ values, because $\reBoundl\simeq\BrooGLSSl$ for $\alpha~=~96^o$, this happens for $\Broo$ close to the reBound, where one observes that 3 mirror solutions are lumped together with the true solution. In that case, the higher $\Broo$, the better the measurement of~$\alpha$. With \sBelle, or with the world average values, the worst case scenario is avoided because the GLSS bound is well below the reBound.
However, nasty situations where some mirror solutions collide with~$\alpha$ can still occur. With \sBelle\ values, this happens by accident (i.e. for $\Broo\neq\BrooGLSSl$) for $\Broo\simeq 2.4 \times 10^{-6}$.

\subsection{Guessestimates for high luminosity ($10~\invab$)}
\label{sec:guesses}
\begin{table}[ht]
\begin{center}
\begin{tabular}{|c|c|c|c||}
\hline
      & $\Brpm~(10^{-6})$ & $\Brpo~(10^{-6})$ &$\Broo~(10^{-6})$ \\ \hline\hline   
$10~\invab$   & $4.55\pm 0.10$   & $5.18\pm 0.18\ $  &$1.9\pm 0.11$ \\ \hline 
$1000~\invfb$ & $4.55\pm 0.19$   & $5.18\pm 0.32\ $  &$1.9\pm 0.22$ \\ \hline
\hline
    & $\Cpm$          & $\Spm$            &$\Coo$     \\ \hline\hline   
$10~\invab$   & $-0.46\pm 0.04$ & $-0.73\pm 0.03$   &$0.8/-0.38 \pm 0.08$ \\ \hline 
$1000~\invfb$ & $-0.46\pm 0.07$ & $-0.73\pm 0.07$   &$0.8/-0.38 \pm 0.26$ \\ \hline
\hline
\end{tabular}
\end{center}
\caption{\it Input values to the fits for the branching ratios, the CP violating asymmetries and their extrapolated errors for $10$ and $1~\invab$. The central values are the W.A. ones. The two values quoted for $\Coo$ are the ones obtained for $\cos(2\alphaeff)\cos(2\alpha_0)$ positive.}
\label{tab:highL}
\end{table}
The value $\alpha~=~96^o$ is chosen together with $\cos(2\alphaeff)$ negative (Section~\ref{sec:toughBoundsection}). The central values for W.A. are used as inputs and the errors are extrapolated to $10~\invab$ (Table~\ref{tab:highL}). Such a luminosity gives a number of $\pizpiz$ events close to $5000$. This set of parameters is consistent with two different values of $\Coo$ as shown Fig.~\ref{fig:cooboo}, which are equal to $0.80$ and $-0.38$. The estimated errors for both $\Coo$ are taken to be $0.08$.

\begin{figure}[t]
\begin{center}
\mbox{\epsfig{file=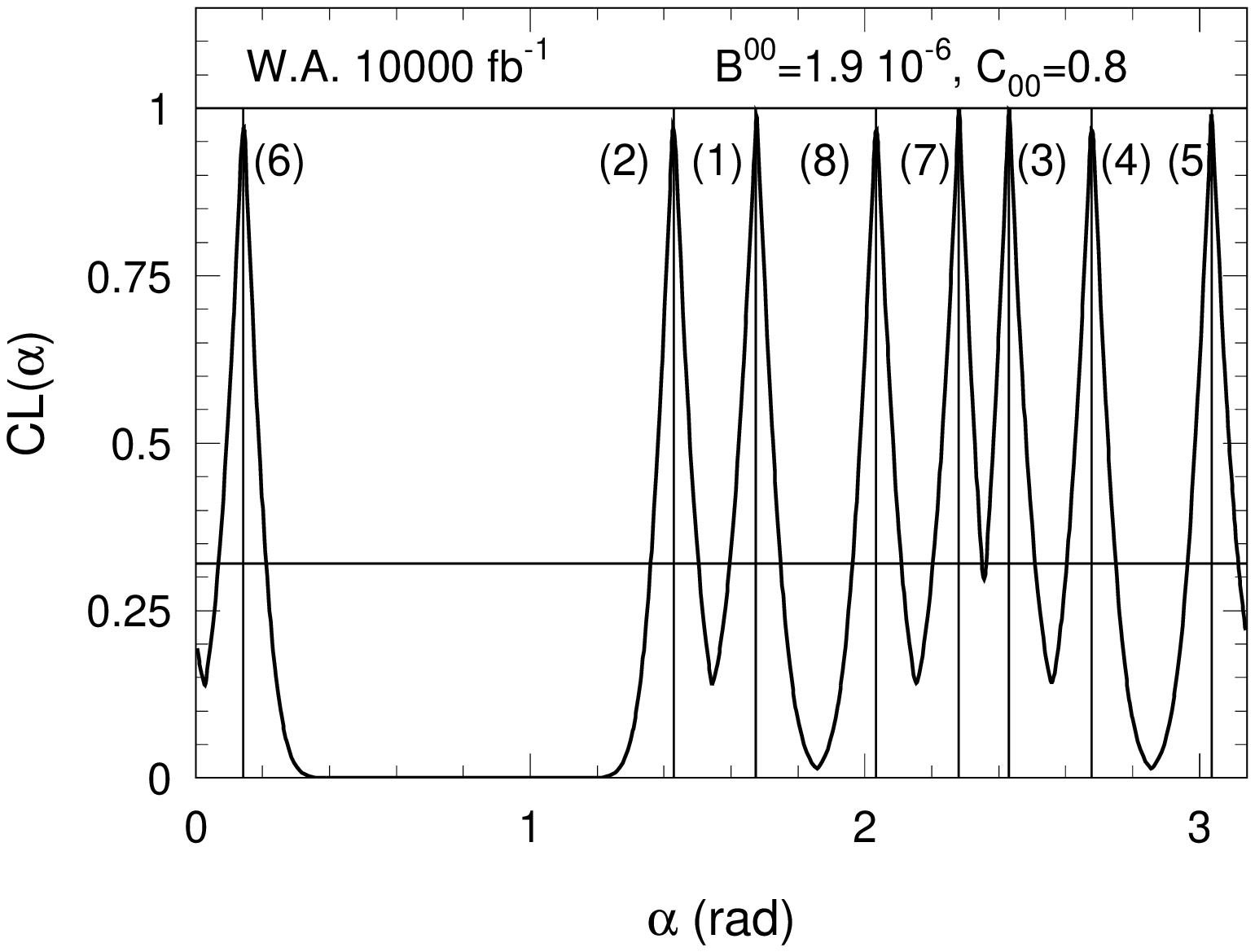,width=0.5\linewidth}
\epsfig{file=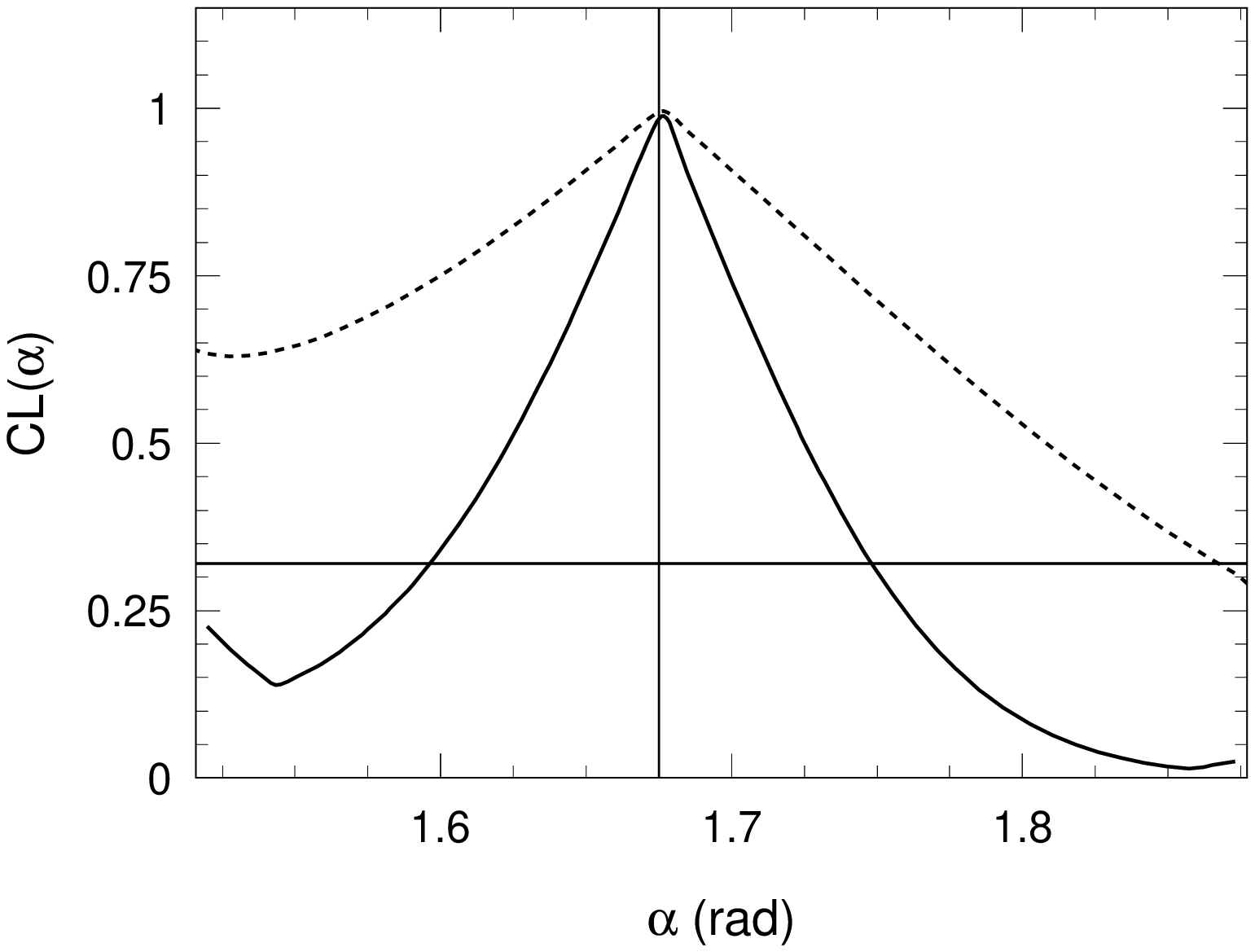,width=0.5\linewidth}}
\end{center}
\caption{\it Confidence Level CL($\alpha$)=Prob$(\chi^2_{\rm min},1)$ obtained for W.A. central values at a luminosity of $10~\invab$ when performing a direct fit with a fine-grain exploration of the parameter space (left). The value chosen for $\Coo$ is $0.8$. One observes the 8 mirror solutions showing up as 8 CL=1 peaks. The right plot is a zoom around the true solution $\alpha~=~96^o$ ($1.67~\rad$). The presence on the left hand side of the nearby mirror solution (2) makes the CL curve asymmetrical. The 32\% CL is reached at $\alpha~=~{1.67}\pm\sigma_\pm$ with $\sigma_+=0.07$ and $\sigma_-=0.08$ (radians). The doted line represents CL($\alpha$) for a luminosity of $1000~\invfb$. In that case, one practically looses all constraints on the left hand side: the mirror solution (2) is too close to be resolved. However, the error on the right hand side remains useful: $\sigma_-=0.19$.}
\label{fig:alphascan}
\end{figure}
\begin{figure}[t]
\begin{center}
\mbox{\epsfig{file=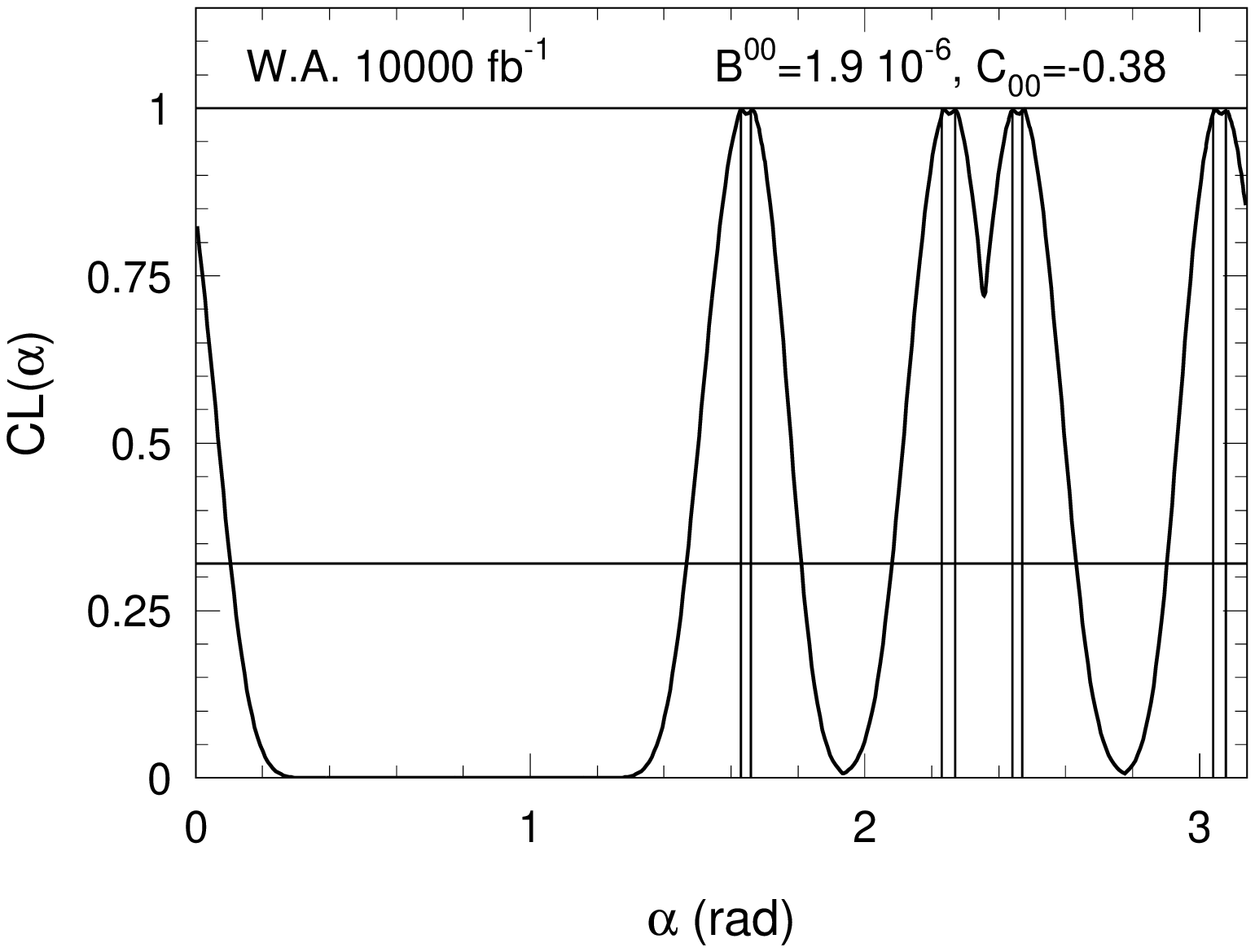,width=0.5\linewidth}
\epsfig{file=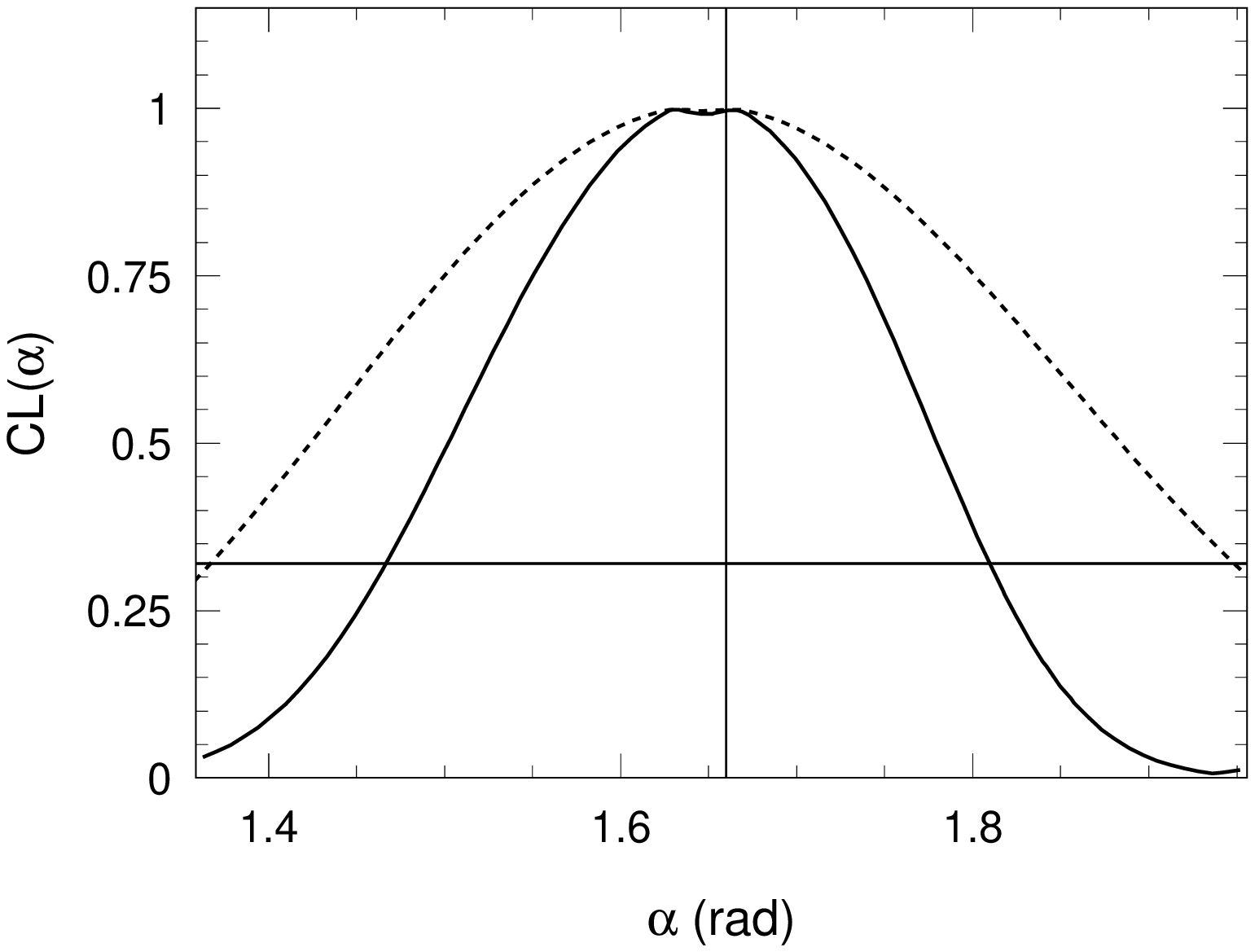,width=0.5\linewidth}}
\end{center}
\caption{\it Confidence Level CL($\alpha$)=Prob$(\chi^2_{\rm min},1)$ obtained for W.A. central values at a luminosity of $10~\invab$ when performing a direct fit with a fine-grain exploration of the parameter space. The value chosen for $\Coo$ is $-0.38$. The CL curves obtained in that case are to be compared with the ones if Fig.~\ref{fig:alphascan} ($\Coo=0.8$). In the present case, the peaks merged partly and are much less pronounced.}
\label{fig:alphascan2}
\end{figure}

The result of the fit is shown on Fig.~\ref{fig:alphascan} ($\Coo=0.8$) and Fig.~\ref{fig:alphascan2} ($\Coo=-0.38$). Six of the mirror solutions appear above $\alpha~=~96^o$. The domain allowed at more than 5\% CL is quite large. Because of that one may be tempted to conclude that $\pi\pi$ does not bring much to CKM metrology. However, one should keep in mind that CKM fits receive contributions from numerous measurements. It is thus misleading to compare $\alpha(\pi\pi)$ to all the other measurements taken at once. Furthermore, precisely because one expects the domain in~$\alpha$ allowed by a global fit to be quite small in the high luminosity area, the low CL values are not anymore relevant. For metrology, what matters then is the sharpness of the 8 peaks. If no mirror solution shows up in the vicinity of the correct~$\alpha$ value, then $\pi\pi$ may contribute to CKM metrology in a very significant way by helping to pinpoint precisely~$\alpha$. This is the case for $\Coo=0.8$ where the 8 peaks are sharp, but this is not the case for $\Coo=-0.38$ where the peaks are dull.

A zoom around $\alpha~=~96^o$ is shown on Fig.~\ref{fig:alphascan} (right) in plain line. Because of the nearby mirror solution (2), the CL curve is not symmetrical. Defining $\sigma_\pm$ by CL$(1.67\pm\sigma_\pm)=32\%$ one gets $\sigma_+=0.07$ and $\sigma_-=0.08$. The negative error blows up fast with decreasing luminosity, because the fit is quickly not able anymore to resolve the true solution from the mirror solution (2). The positive error scales almost with the square root of the luminosity because the CL is almost at zero down the pic.
The CL$(\alpha)$ for a luminosity of $1000~\invfb$ is shown as well in dotted line. For this case the positive error $\sigma_-=0.19$ is slightly smaller than the (ill-defined) actual precision on~$\alpha$ from the global CKM fit ($=0.23$). Obviously, before then, the measurement of $\Delta m_s$ by the Tevatron experiments will have changed the global CKM picture. 

\subsection{Two scenarii at $500~\invfb$ for \babar}
\label{sec:TheNastyScenarioSection}
\begin{table}[ht]
\begin{center}
\begin{tabular}{|c|c|c|c||}
\hline
      & $\Brpm~(10^{-6})$ & $\Brpo~(10^{-6})$ &$\Broo~(10^{-6})$ \\ \hline\hline   
scenario I  & $4.7\pm 0.26$   & $5.5\pm 0.43\ $  &$0.7\pm 0.29$ \\ \hline 
scenario II & $4.7\pm 0.26$   & $5.5\pm 0.43\ $  &$1.7\pm 0.29$ \\ \hline
\hline
    & $\Cpm$          & $\Spm$            &$\Coo$     \\ \hline\hline   
scenario I  & $-0.19\pm 0.09$ & $-0.40\pm 0.10$   &$0.47\pm 1.29$ \\ \hline 
scenario II & $-0.19\pm 0.09$ & $-0.40\pm 0.10$   &$0.29\pm 0.58$ \\ \hline
\hline
\end{tabular}
\end{center}
\caption{\it Input values to the fits for the branching ratios, the CP violating asymmetries and their extrapolated errors. Scenarii I and II use the \babar\ values for $\Brpm$, $\Brpo$, $\Cpm$ and $\Spm$ and two distinct $\Broo$ guesses.}
\label{tab:scenar1et2}
\end{table}
Now, only the \babar\ central values are considered, at a luminosity of $500~\invfb$ which is planned to be reached in 2006 by \babar. Considering the upper left plot on Fig.~\ref{fig:switchyard}, one considers two scenarii for $\Broo$. The worst case scenarii are obtained when several mirror solutions gather around the true solution, staying about one $\sigma[\alpha]$ away. With \babar\ values, this happens for $\Broo$ close to the reBound value $\reBoundl = 0.6 \times 10^{-6}$, where 3 mirror solutions are lumped together with the true solution. As discussed in Section~\ref{sec:closeformforalpha} this happens because $\alphaeff$ being close to~$\alpha$, the reBound is close to the GLSS bound where the mirror solutions are exactly degenerate. To show the effect, one considers a scenario with the barely acceptable value $\Broo=0.7 \times 10^{-6}$: this is the scenario I. Scenario II is one where, taking the opposite stand, $\Broo$ is far from the reBound. The degeneracy between the mirror solutions is then lifted which lets presume that the measurement of $\alpha$ will be easier. The value $\Broo=1.7 \times 10^{-6}$ is chosen for this case.

\begin{figure}[t]
\begin{center}
\mbox{\epsfig{file=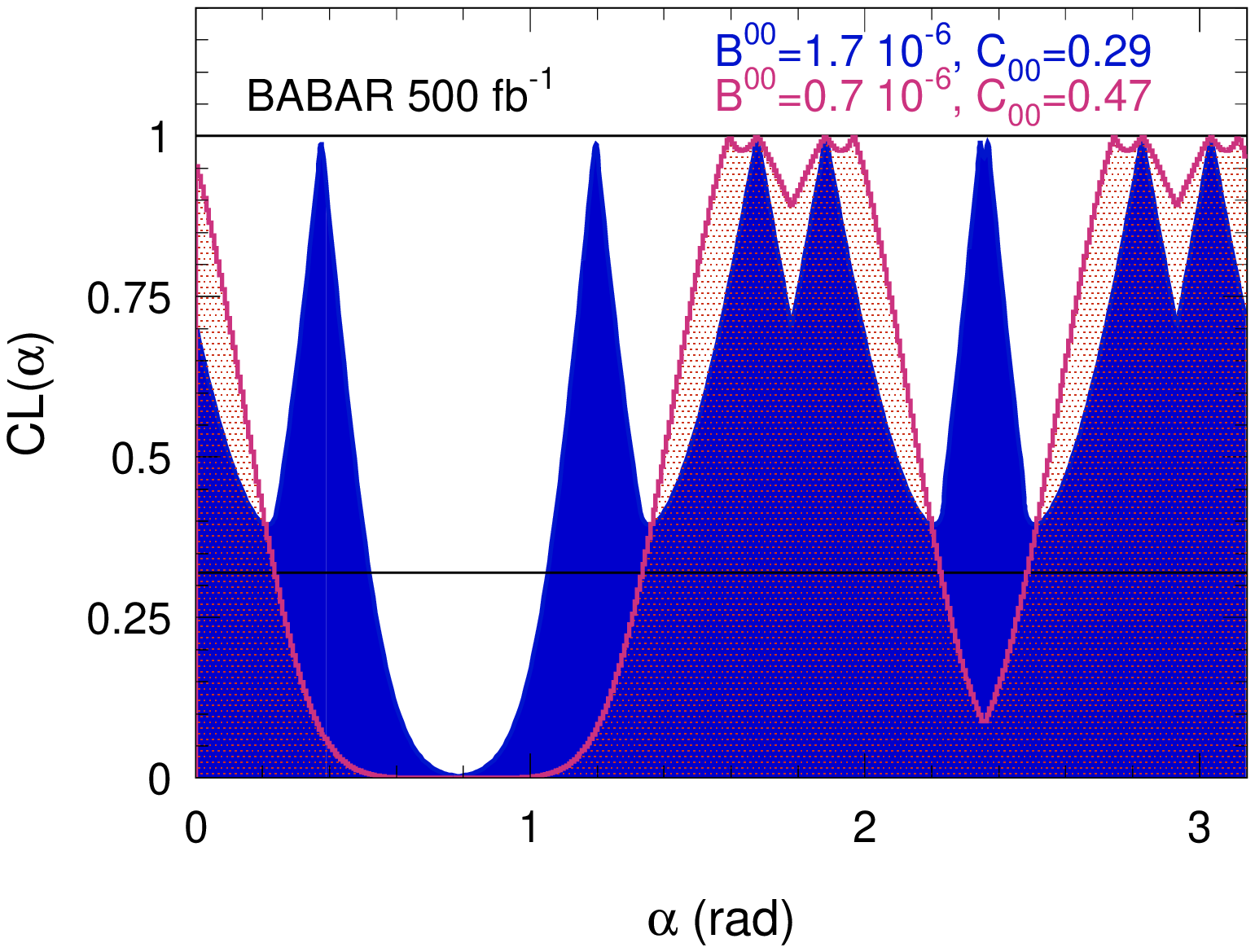,width=0.5\linewidth}
\epsfig{file=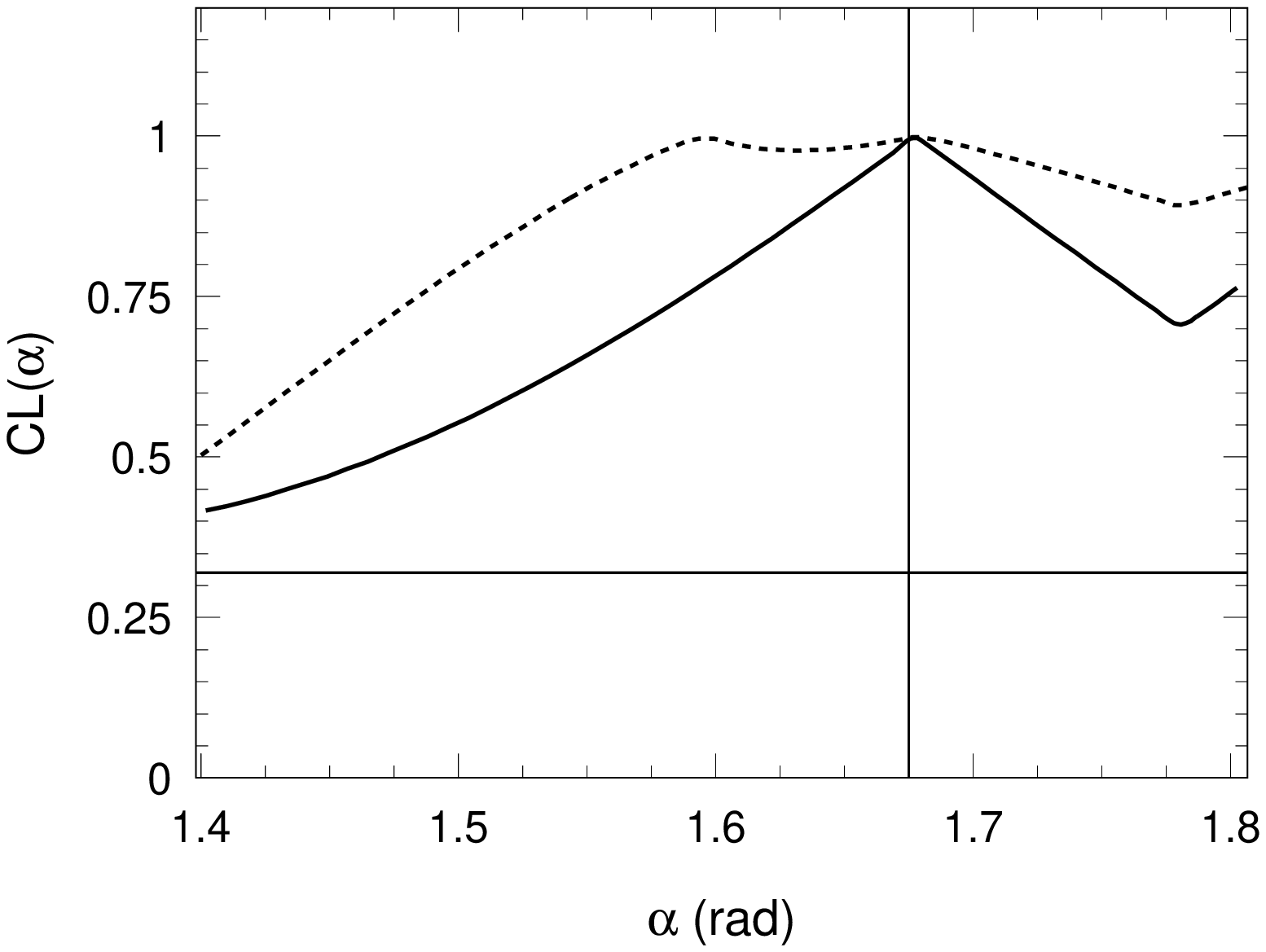,width=0.5\linewidth}}
\end{center}
\caption{\it Confidence Level CL($\alpha$)=Prob($\chi^2_{\rm min},1)$ obtained for scenario I and scenario II. The luminosity is $500~\invfb$ and $\alpha~=~96^o$. The right plot is a zoom around the value $\alpha~=~96^o$ ($1.67~\rad$) for $\Broo=0.7 \times 10^{-6}$ (dotted line) and $\Broo=1.7 \times 10^{-6}$ (plain line).}
\label{fig:alphascannasty}
\end{figure}

The values used for these two scenarii are presented in Table~\ref{tab:scenar1et2}. The result of the fits are shown Fig.~\ref{fig:alphascannasty}. A zoom of the fits around the true value of $\alpha$ is displayed on the right.

Whereas, in scenario II, the larger value of the branching ratio $\Broo$ leaves some power to the analysis, in scenario I, the vicinity of the mirror solutions deprives the analysis of most of its interest.

\section{Standard Model Tests: the $(\Broo,\Coo)$ plot}
\label{sec:StandarModelstestsection}
The above discussion offers a perspective on the future capabilities of $\pi\pi$ analysis to constraint the angle~$\alpha$: everything can happen in the few coming years, including a precise measurement of $\alpha$ from the $B \to \pi\pi$ decays.

Testing the Standard Model is a different topic. Although this question can be addressed by comparing the constraints on $\alpha$ from the standard CKM fit to the ones obtained from $B \to \pi\pi$ decays, it is a clumsy way to proceed. Indeed, one has to deal with the mirror solutions which are irrelevant when one is only willing to know whether or not the Standard Model can accomodate the set of six measurements $\Brpm$, $\Brpo$, $\Broo$, $\Cpipi$, $\Spipi$ and $\Coo$ knowing $\alpha$ from other measurements. Since the last precise measurements are going to be $\Broo$ and $\Coo$ (the latter being still out of reach) it is appropriate to probe the Standard Model by considering the $(\Broo,\Coo)$ plane.

\begin{figure}[t]
\begin{center}
\mbox{\psfig{file=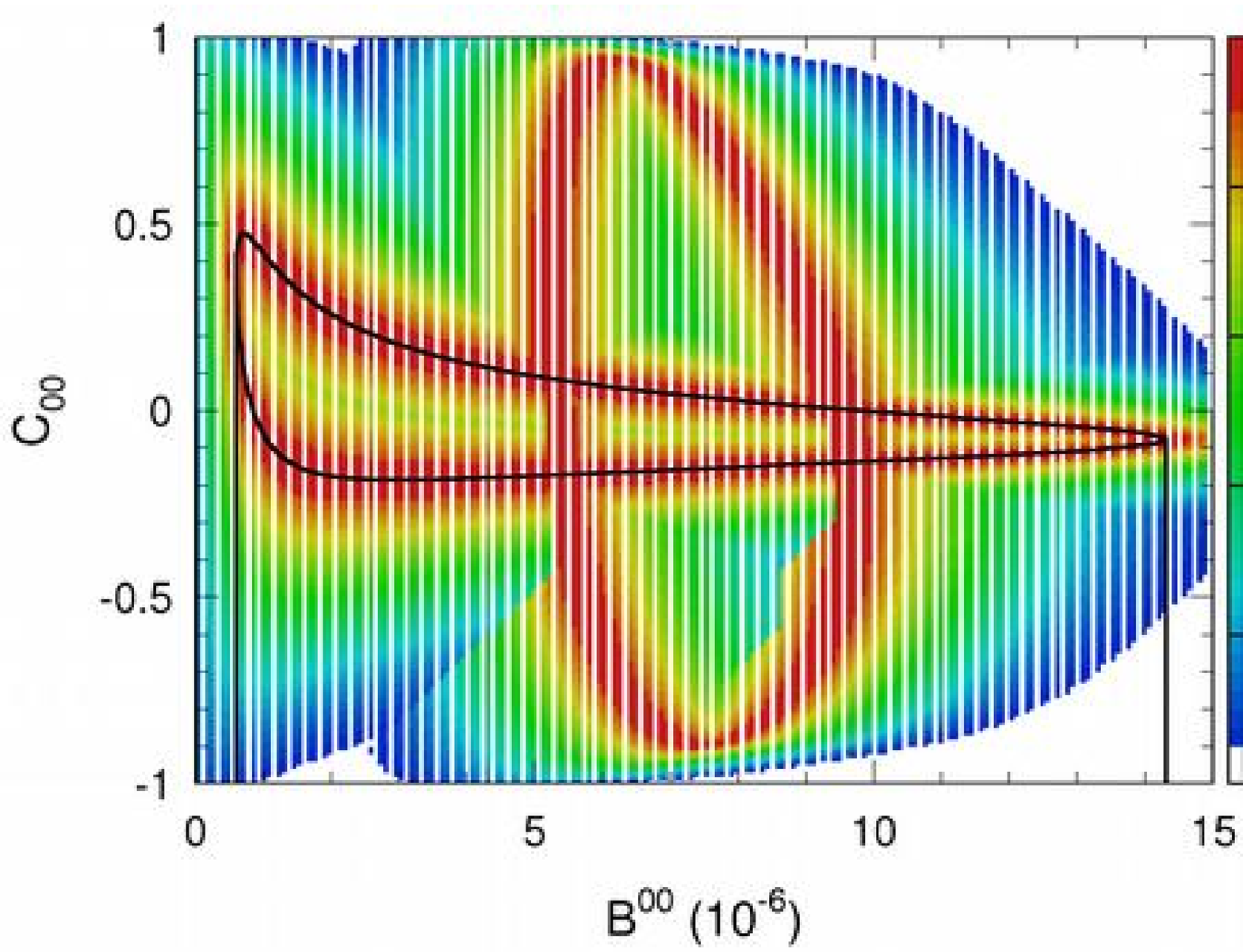,width=0.34\linewidth}
\psfig{file=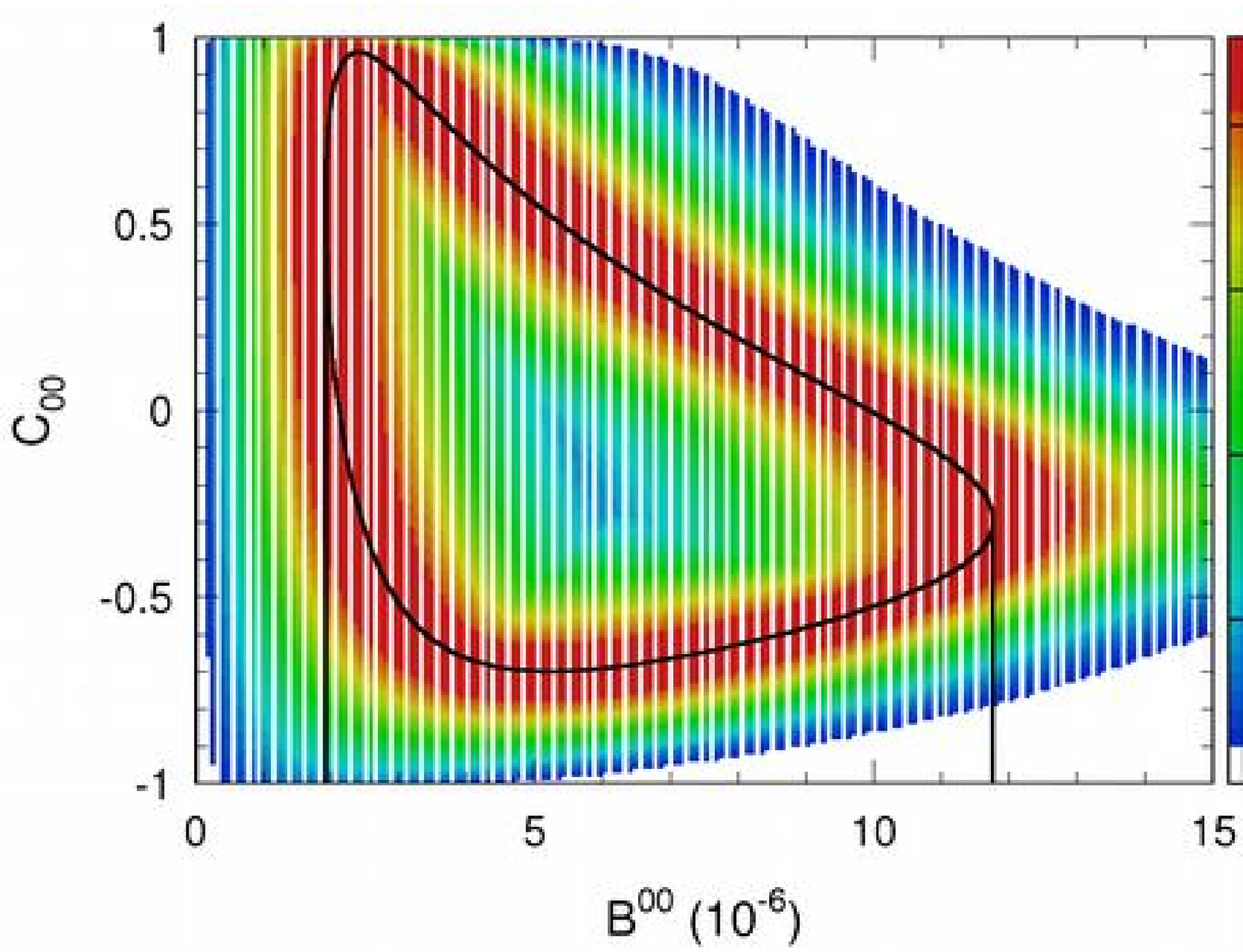,width=0.34\linewidth}
\psfig{file=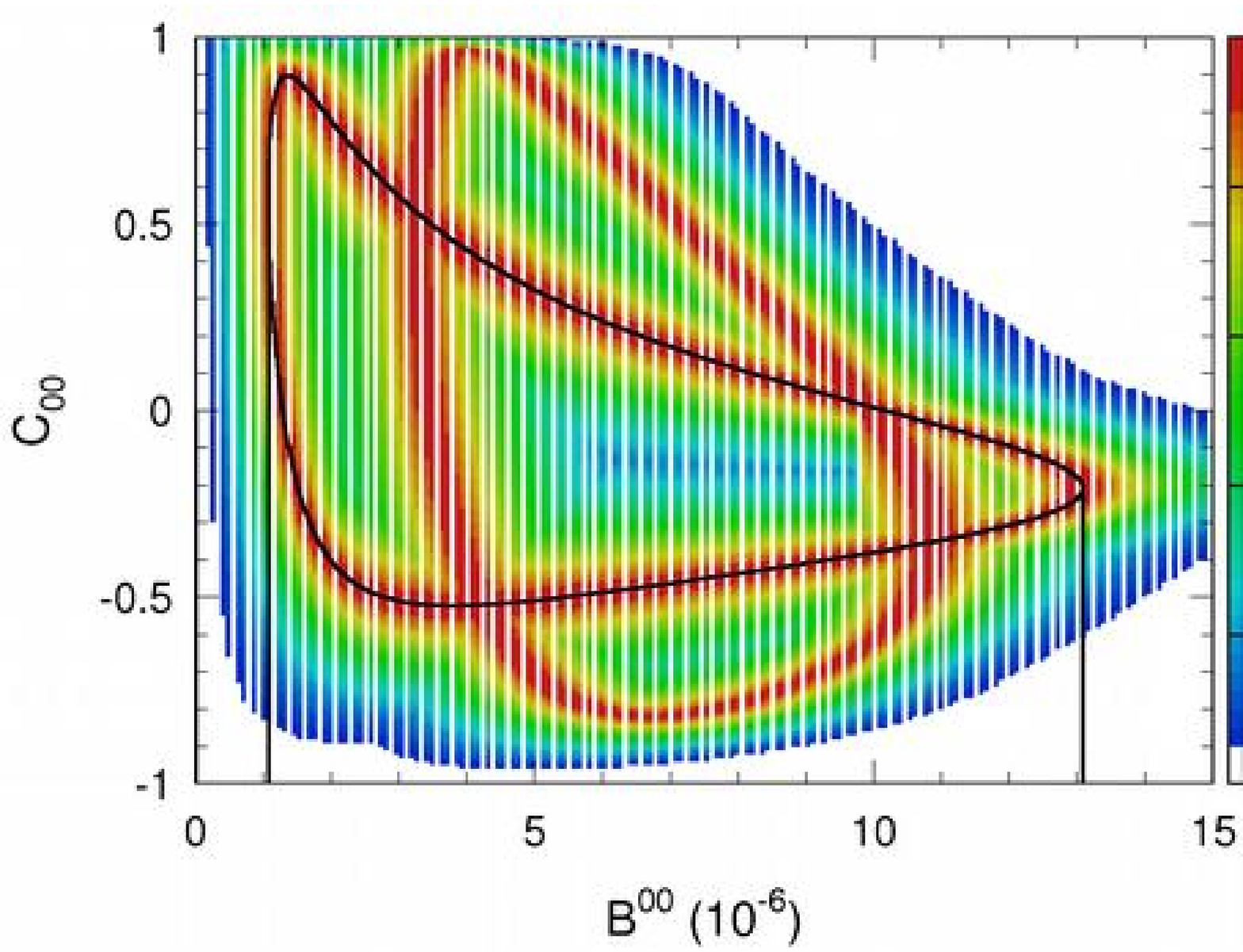,width=0.34\linewidth}}
\caption{\it The $(\Broo,\Coo)$ plane is presented for \babar\ (left), \sBelle\ (middle) and W.A. (right), using the actual luminosity and $\alpha~=~96^o$. The curves superimposed are the ones given by Eq.(\ref{eq:master}) for $\cos(2\alphaeff)$ negative (see Fig.~\ref{fig:cooboo}).}
\label{fig:ckmpipi_am}
\end{center}
\end{figure}

The $\Coo(\Broo)$ function introduced in Section~\ref{sec:coo} has shown the possible contour, for \babar, \sBelle\ and W.A., of $\Coo$ as a function of $\Broo$ (Fig.~\ref{fig:cooboo}). Here is now presented the plane ($\Broo,\Coo$) when a fit is performed using $\Brpm$, $\Brpo$, $\Cpipi$, $\Spipi$ and $\alpha$, in order to evaluate the CL at each point. To look at this plane is a proper way to handle data because no bound is needed, nor dealing with mirror solutions. The results for the three experimental cases are presented on Fig.~\ref{fig:ckmpipi_am} for the present luminosity and $\alpha~=~96^o$ with the curves of Fig.~\ref{fig:cooboo} superimposed. Considering the allowed region ($\Broo<4 \times 10^{-6}$), one observes that there is practically no constraints on $\Coo$, neither for \babar, \sBelle\ or W.A. Similarly, with the present level of knowledge, any value of $\Broo$ can be accomodated by the Standard Model, except exceedingly small ones.

\begin{figure}
\begin{center}
\mbox{\psfig{file=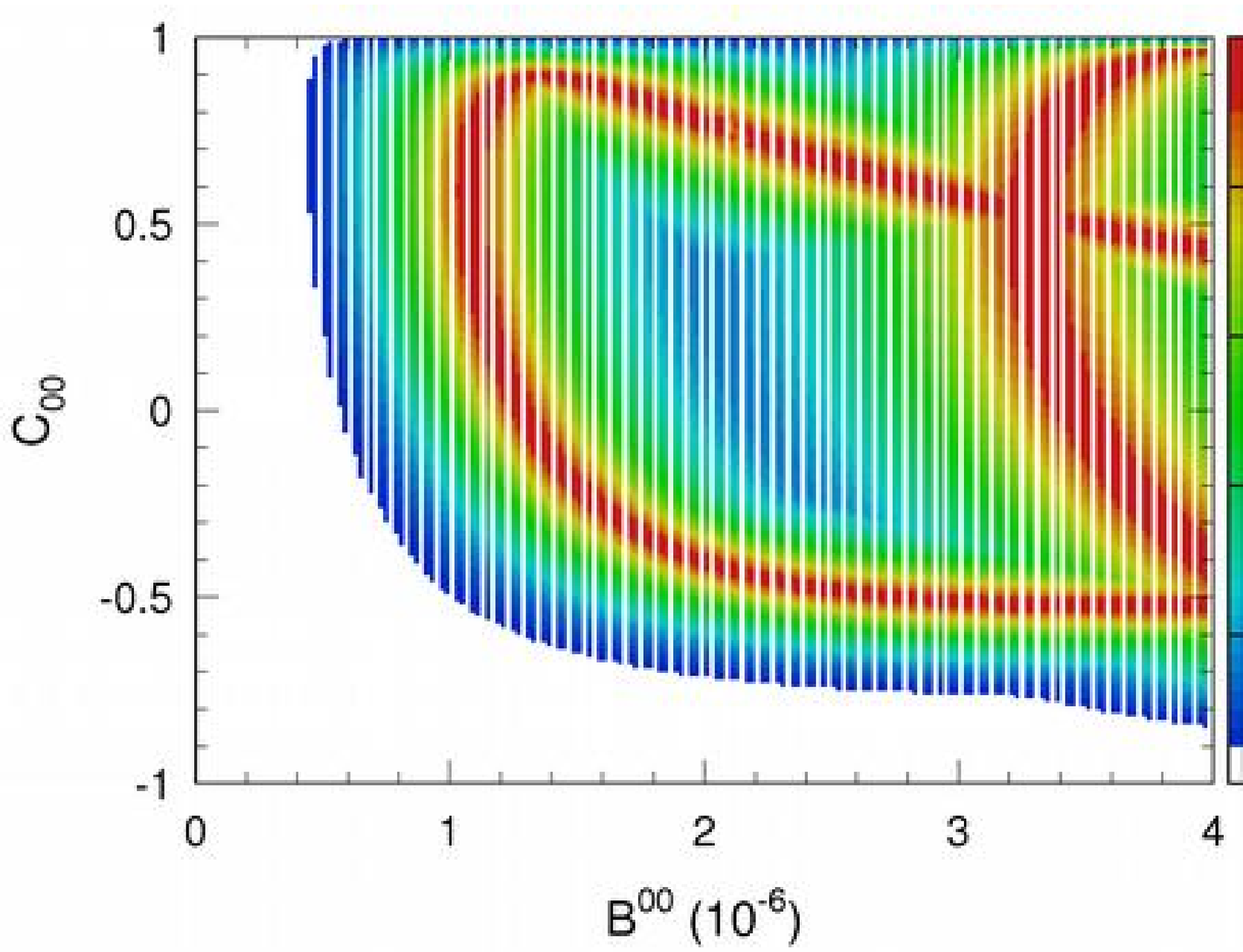,width=0.44\linewidth}
\psfig{file=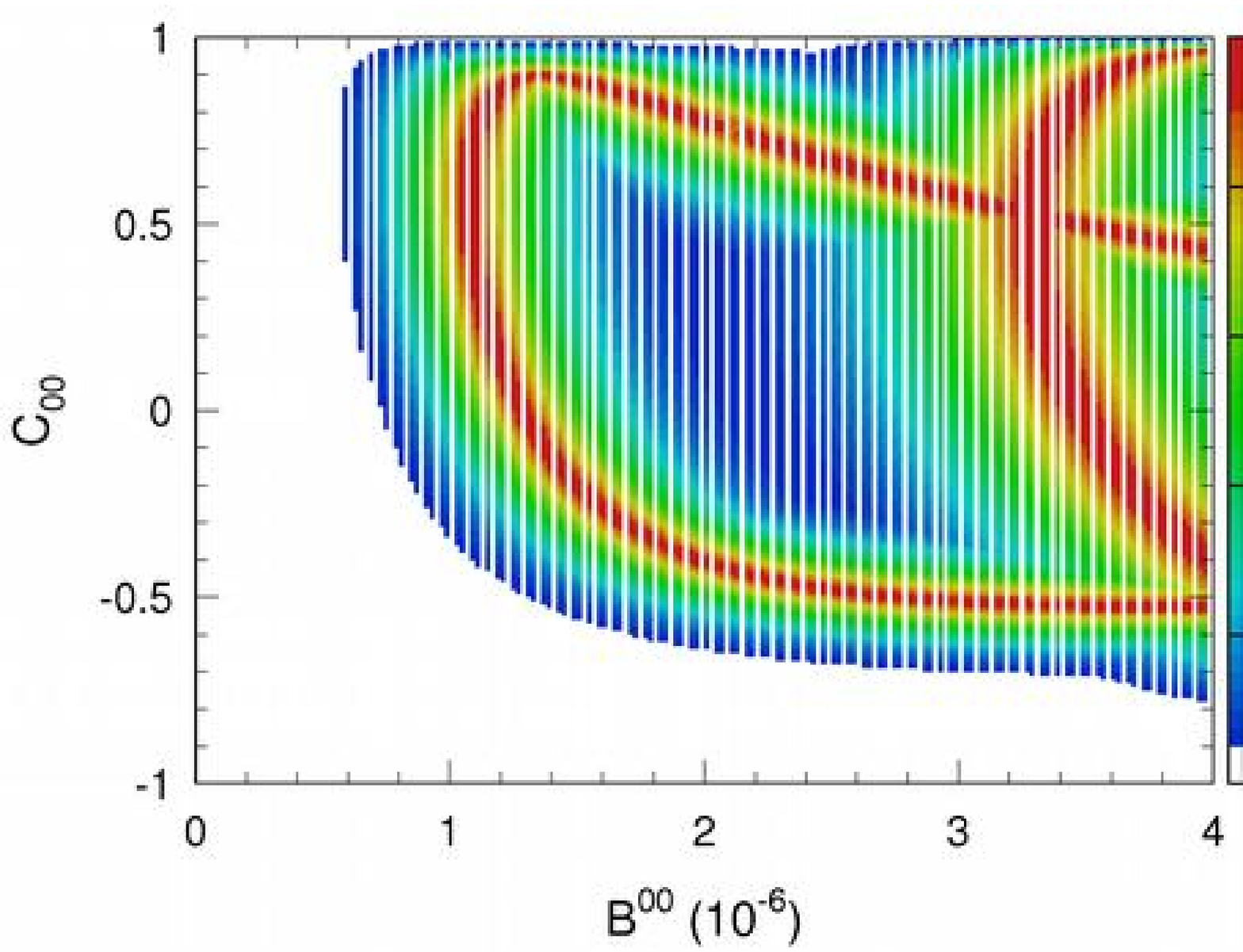,width=0.44\linewidth}}
\caption{\it The $(\Broo,\Coo)$ plane is presented using the world average branching ratios and CP parameters, for $500~\invfb$ (left) and $1000~\invfb$ (right).}
\label{fig:ckmpipi_am_WA}
\end{center}
\end{figure}

Figure~\ref{fig:ckmpipi_am_WA} represent the plane ($\Broo,\Coo$) considering the W.A. results, for a luminosity of $500$ and $1000~\invfb$. To rule out the Standard Model would imply an accurate measurement of a large negative value of $\Coo$ and/or a very small value of $\Broo$.

\section{Conclusion}
The constraints on the angle~$\alpha$ are revisited, considering the three branching ratios $\Brpm$, $\Brpo$ and $\Broo$, and the three CP parameters $\Cpipi$, $\Spipi$ and $\Coo$. A new bound is established on $\Broo$ taking advantage of the constraints on~$\alpha$ available from the standard CKM fit. The function $\Coo(\Broo)$ is introduced and the use of the $(\Broo,\Coo)$ plane is advocated as a means to probe new physics, when the measurement of $\Coo$ will be available. The Grossman-Quinn bound is discussed and a compact formula is provided which highlights its domain of validity.

When considering \babar\ results, three cases can be considered: 1) if $\Broo$ reaches exactly its lower bound, there is no ambiguity for $\alpha$ (but this is academic), 2) if $\Broo$ is only close to its lower bound, the locations of the mirror solutions are varying rapidly as a function of $\Broo$ and the evaluation of~$\alpha$ becomes difficult, 3) if $\Broo$ is far enough from the lower bound, the degeneracy between the mirror solutions tend to be lifted and the measurement of $\alpha$ may become feasible with only a medium statistics. The current period is very exciting, anything can happen in the next three years: including a rather precise determination of~$\alpha$.

The introduction of the ($\Broo,\Coo$) plane in order to test the Standard Model shows that it will be difficult to rule out this latter.


\end{document}

%% file: definitions.tex
\def\babar{\mbox{\slshape B\kern-0.1em{A}\kern-0.1em
    B\kern-0.1em{A\kern-0.2em R}}}

\def\sPlot{\hbox{$_s$}{\cal P}lot}

\def\alphaeff{\alpha_{\rm eff}}
\def\phibar{\overline{\phi}}
\def\Apm{{\cal A}^{+-}}
\def\Abarpm{\overline{{\cal A}}^{+-}}
\def\Apo{{\cal A}^{+0}}
\def\Abarpo{{\cal A}^{-0}}

\def\Aoo{{\cal A}^{00}}
\def\Abaroo{\overline{{\cal A}}^{00}}
\def\a{a}
\def\abar{\bar{\a}}
\def\cc{c}
\def\ccbar{\bar{\cc}}
\def\ss{s}
\def\ssbar{\bar{\ss}}
\def\pio{{\pi\over 2}}

\def\Br{{\cal B}}
\def\Brpm{\Br^{+-}}
\def\Brpo{\Br^{+0}}
\def\Broo{\Br^{00}}
\def\Cpm{C_{\pi\pi}}
\def\Spm{S_{\pi\pi}}
\def\Coo{C_{00}}

\def\BrooGLSSu{\Br^{00}_{\rm GLSS:u}}
\def\BrooGLSSl{\Br^{00}_{\rm GLSS:l}}

\def\Soo{S_{00}}
\def\sBelle{Belle$^\ast$}
\def\CL{{\rm CL}}
\def\hyp{{\rm hyp}}
\def\exp{{\rm exp}}
\def\best{{\rm best}}
\def\Cpmhyp{C_{\pi\pi}^\hyp}
\def\Spmhyp{S_{\pi\pi}^\hyp}
\def\Cooexp{C_{00}^\exp}
\def\Cpmexp{C_{\pi\pi}^\exp}
\def\Spmexp{S_{\pi\pi}^\exp}
\def\Brpmexp{\Br^{+-}_\exp}
\def\Brpoexp{\Br^{+0}_\exp}
\def\Brooexp{\Br^{00}_\exp}
\def\Cpmbest{C_{\pi\pi}^\best}
\def\Spmbest{S_{\pi\pi}^\best}
\def\sinmax{S_{\rm GQ}}

\def\newBoundl{\Br^{00}_{\alpha:{\rm l}\pm}}
\def\newBoundu{\Br^{00}_{\alpha:{\rm u}\pm}}

\def\reBoundl{\Br^{00}_{\alpha:{\rm l}+}}

\def\toughBoundl{\Br^{00}_{\alpha:{\rm l}-}}

\def\reBoundlepscos{\Br^{00}_{\alpha:{\rm l}{\pm}}}
\def\reBounduepscos{\Br^{00}_{\alpha:{\rm u}{\pm}}}

\def\Cpipi{C_{\pi\pi}}
\def\Spipi{S_{\pi\pi}}
\def\EPJ{{\em Eur. Phys. J.}}
\def\PRL{{\em Phys. Rev. Lett.}}
\def\pipi{\pi^+\pi^-}
\def\pippiz{\pi^+\pi^0}
\def\Bztopipi{B^0 \to \pipi}
\def\Bptopippiz{B^+ \to \pippiz}

\def\invfb{\mbox{fb}^{-1}}
\def\invab{\mbox{ab}^{-1}}
\def\ea{{\em et al.}}
\def\pipi  {{\pi^+\pi^-}}
\def\piz   {{\pi^0}}

\def\pizpiz{\pi^0\pi^0}

\def\Bztopizpiz{B^0 \to \pizpiz}
\def\Btopippiz{B^+ \to \pi^+\piz}

\def\ckmfitter{{\em CKMfitter}}

\def\tauBp{\tau{B^{+}}}
\def\tauBo{\tau{B^{0}}}
\def\pep2{PEP-II}
\def\BF{$B$ Factory}
\def\abf {asymmetric \BF}
\def\rad{\rm rad}